\documentclass[twocolumn,american]{article}
\usepackage[T1]{fontenc}
\usepackage{geometry}
\usepackage{color}
\usepackage{units}
\usepackage{textcomp}
\usepackage{amsmath}
\usepackage{amsthm}
\usepackage{amssymb}
\usepackage{graphicx}

\makeatletter
\usepackage{abstract} 
\newcommand{\makeabstract}{\@ifundefined{abstractcontent}{}{\begin{abstract}\abstractcontent\end{abstract}}}
\newcommand{\makefrontmatter}{\if@twocolumn{\twocolumn[\maketitle\makeabstract\vskip2\baselineskip]\saythanks}\else{\maketitle\makeabstract}\fi}

\usepackage{amsmath}
\usepackage[bb=dsserif]{mathalpha}
\usepackage{bm}
\usepackage[latin1]{inputenc}
\usepackage{csquotes}

\makeatother

\usepackage{babel}
\usepackage[bibstyle=numeric,citestyle=numeric-comp-archaeology,sorting=none]{biblatex}
\begin{document}
\title{Attention to Entropic Communication}
\author{Torsten En{\ss}lin$^{1,2,3}$, Carolin Weidinger$^{1,2}$, Philipp
Frank$^{1}$ \\
{\small{}$^{1}$ Max Planck Institute for Astrophysics, Karl-Schwarzschild-Str.
1, 85748 Garching, Germany}\\
{\small{}$^{2}$ Ludwig-Maximilians-Universit\"at M\"unchen, Geschwister-Scholl-Platz
1 80539 Munich, Germany}\\
{\small{}$^{3}$ Excellence cluster ORIGINS, Boltzmannstr. 2 85748
Garching, Germany}}
\newcommand{\abstractcontent}{The concept of attention, numerical weights that emphasize the importance
of particular data, has proven to be very relevant in artificial intelligence.
Relative entropy (RE, aka Kullback-Leibler divergence) plays a central
role in communication theory. Here we combine these concepts, attention
and RE. RE guides optimal encoding of messages in bandwidth-limited
communication as well as optimal message decoding via the maximum
entropy principle (MEP). In the coding scenario, RE can be derived
from four requirements, namely being analytical, local, proper, and
calibrated. Weighted RE, used for attention steering in communications,
turns out to be improper. To see how proper attention communication
can emerge, we analyze a scenario of a message sender who wants to
ensure that the receiver of the message can perform well-informed
actions. If the receiver decodes the message using the MEP, the sender
only needs to know the receiver's utility function to inform optimally,
but not the receiver's initial knowledge state. In case only the curvature
of the utility function maxima are known, it becomes desirable to
accurately communicate an attention function, in this case a by this
curvature weighted and re-normalized probability function. Entropic
attention communication is here proposed as the desired generalization
of entropic communication that permits weighting while being proper,
thereby aiding the design of optimal communication protocols in technical
applications and helping to understand human communication. For example,
our analysis shows how to derive the level of cooperation expected
under misaligned interests of otherwise honest communication partners.\linebreak{}
\textbf{\emph{Key words:}} \emph{information theory; communication
theory; entropy; attention; weighted Kullback Leibler Divergence}}

\makefrontmatter
\vspace{0cm}

\section{Introduction}

\subsection{Relative Entropy}

The work of Shannon \cite{shannon} and Jaynes \cite{PhysRev.106.620,PhysRev.108.171,jaynes1963information,jaynes1968prior,jaynes03}
made it clear that entropy and its generalizations \cite{grunwald2004game,gneiting2007strictly,crupi2018generalized},
in particular to relative entropy \parencite{10.1214/aoms/1177729694},
have a root in communication theory. Already introduced by \textcite{gibbs1906thermodynamics}
in 1906 to thermodynamics, relative entropy plays nowadays a central
role in artificial intelligence \parencite{opper2001advanced,10.7551/mitpress/1100.003.0007,10.7551/mitpress/1100.003.0014,10.7551/mitpress/1100.003.0020,10.7551/mitpress/1100.003.0021},
particularly for variational autoencoder \parencite{kingma2014stochastic,kingma2019introduction,milosevic2021bayesian,zacherl2021probabilistic},
as well as in information field theory \parencite{2010PhRvE..82e1112E,knollmuller2019metric,frank2021geometric}.
Relative entropy permits to code maximal informative messages and
the Maximum Entropy Principle (MEP) \cite{PhysRev.106.620,PhysRev.108.171,jaynes1963information,jaynes1968prior,jaynes03}
to decode such, as detailed in the following.

The \textbf{relative entropy }between two probability densities $\mathcal{P}(s|I_{\text{A}})$
and $\mathcal{P}(s|I_{\text{B}})$ on an unknown signal or situation
$s\in\mathcal{S}$

\begin{eqnarray}
\mathcal{D}_{s}(I_{\text{A}},I_{\text{B}}) & := & \int_{\mathcal{S}}ds\,\mathcal{P}(s|I_{\text{A}})\,\ln\frac{\mathcal{P}(s|I_{\text{A}})}{\mathcal{P}(s|I_{\text{B}})}\label{eq:relative-entropy}
\end{eqnarray}
measures the amount of information in nits lost by degrading some
knowledge $I_{\text{A}}$ to knowledge $I_{\text{B}}$. The letters
A and B stand for Alice and Bob, who are communication partners. Alice
can use the relative entropy to decide which message she wants to
send to Bob in order to inform him best. $I_{\text{A}}$ is Alice's
background information before and after her communication.

We assume that Alice knows how to communicate such that Bob updates
his previous knowledge state $I_{0}$ to $I_{\text{B}}$.\footnote{In this paper we chose to take the \emph{objective Bayesian} perspective
\cite{williamson2010defence,landes2013objective,landes2015probabilism},
in that we assume that $\mathcal{P}(s|I)$ is uniquely defined by
$I$ and thus that $\mathcal{P}(s|I_{\text{A}})=\mathcal{P}(s|I_{\text{B}})$
for all $s\in\mathcal{S}$ implies that $I_{\text{A}}$ and $I_{\text{B}}$
carry equivalent information on $s$. This choice is convenient, since
simplifying the discussion, but not essential.}

The functional form of the relative entropy can be derived from various
lines of argumentation \cite{Bernardo,caticha2006updating,knuth2012foundations,2017Entrp..19..402L,Harremoes,gkelsinis2020theoretical}.
As the most natural information measure relative entropy plays a central
role in information theory. It is often used as the quantity to be
minimized when deciding which of the possible messages shall be sent
through a bandwidth limited communication channel that does not permit
the full transfer of $I_{\text{A}}$, but also in other circumstances.

As we will discuss in more detail, relative entropy as specified by
Eq.\ \ref{eq:relative-entropy} is uniquely determined up to a multiplicative
factor as the measure to determine the optimal message to be sent
under the requirements of it being analytical (all derivatives w.r.t.\ to
the parameters in $I_{\text{B}}$ exist everywhere), local (only the
$s$ that happens will matter in the end), proper (to favor $I_{\text{B}}=I_{\text{A}}$)
\cite{mccarthy1956measures,winkler1968good,gneiting2007strictly},
and calibrated (being zero when $I_{\text{B}}=I_{\text{A}}$). Our
derivation is a slight modification of that given in \textcite{2017Entrp..19..402L}.

A number of attempts have been made to introduce weights into the
relative entropy \cite{di2007weighted,yasaei2013kullback,gkelsinis2020theoretical,zhou2021application,doi:10.1080/03610918.2022.2108053,YE202229,10.2307/30045207}.
Some of these go back to \textcite{GUIASU198663,1054185}. Most of
them can be summarized by the \textbf{weighted relative entropy},

\begin{eqnarray}
\mathcal{\widetilde{D}}_{s}^{(w)}(I_{\text{A}},I_{\text{B}}) & := & \int_{\mathcal{S}}ds\,w(s)\,\mathcal{P}(s|I_{\text{A}})\,\ln\frac{\mathcal{P}(s|I_{\text{A}})}{\mathcal{P}(s|I_{\text{B}})},\label{eq:weighted-relative-entropy}
\end{eqnarray}
with weights $w(s)$, for which $w(s)\ge0$ holds for all $s\in\mathcal{S}$.

The extension of relative entropy to weighted relative entropy, as
given by Eq.\ \ref{eq:weighted-relative-entropy}, appears to be
attractive, as it can reflect scenarios in which not all possibilities
in $\mathcal{S}$ are equally important. For example, detailed knowledge
on the subset of situations $\mathcal{S}_{\text{B dead}}\subset\mathcal{S}$
in which Bob's decisions do not matter to him are not very relevant
for the communication, as he can not gain much practical use from
it. Therefore, Alice should not waste valuable communication bandwidth
for details within $\mathcal{S}_{\text{B dead}}$, but use it to inform
him about the remaining situations $\mathcal{S}_{\text{B alive}}=\mathcal{S}\backslash\mathcal{S}_{\text{B dead}}$
for which being well informed makes a difference to Bob. However,
despite being well motivated, weighted relative entropy is not proper
in the mathematical sense for non-constant weighting functions, as
we will show and was already recognized before \cite{gkelsinis2020theoretical}.

\subsection{Relative Attention Entropy}

Given the success of attention weighting in artificial intelligence
research \parencite{chen2015abc,vaswani2017attention,lindsay2020attention},
in particular in transformer networks \parencite{lindsay2020attention},
the question arises whether a form of weighted relative entropy exists
that is proper. In order to understand how the weighting should be
included we investigate a specific communication scenario, in which
weighted and re-normalized probabilities of the form
\begin{equation}
\mathcal{A}^{(w)}(s|I):=\frac{w(s)\,\mathcal{P}(s|I)}{\int_{\mathcal{S}}ds'\,w(s')\,\mathcal{P}(s'|I)}\label{eq:attention}
\end{equation}
appear naturally as the central element of communication. We will
call a quantity of this form \textbf{attention function}, \textbf{attention
density function},\textbf{ }or briefly \textbf{attention}\footnote{The term ``attention'' seems appropriate for this quantity: \emph{Attention
is the concentration of awareness on some phenomenon to the exclusion
of other stimuli.\parencite{EncBrit-Attention} It is a process of
selectively concentrating on a discrete aspect of information, whether
considered subjective or objective} \cite{enwiki:1187259873}. If
``awareness on some phenomenon'' can be read as the probability
associated with it, the weighting done in our construction of attention
then concentrates the awareness on relevant information.} when there is no risk of confusion with other attention concepts
(as within this paper).

The corresponding \textbf{relative attention entropy} 
\begin{eqnarray}
\mathcal{D}_{s}^{(w)}(I_{\text{A}},I_{\text{B}}) & \negmedspace\!:=\negmedspace\! & \int_{\mathcal{S}}ds\,\mathcal{A}^{(w)}(s|I_{\text{A}})\,\ln\frac{\mathcal{A}^{(w)}(s|I_{\text{A}})}{\mathcal{A}^{(w)}(s|I_{\text{B}})},\label{eq:relative-attention-entropy}
\end{eqnarray}
leads to proper communication, in case $w(s)>0$ for all $s\in\mathcal{S}$,
as we show in the following. The minimum of the relative attention
entropy is given by $\mathcal{A}^{(w)}(s|I_{\text{B}})=\mathcal{A}^{(w)}(s|I_{\text{A}})$,
which in case $w(s)>0$ for all $s\in\mathcal{S}$ implies $\mathcal{P}(s|I_{\text{B}})=\mathcal{P}(s|I_{\text{A}})$
since 
\begin{eqnarray}
\mathcal{P}(s|I_{\text{B}}) & = & \frac{\mathcal{A}^{(w)}(s|I_{\text{B}})/w(s)}{\int_{\mathcal{S}}ds'\,\mathcal{A}^{(w)}(s'|I_{\text{B}})/w(s')}\nonumber \\
 & = & \frac{\mathcal{A}^{(w)}(s|I_{\text{A}})/w(s)}{\int_{\mathcal{S}}ds'\,\mathcal{A}^{(w)}(s'|I_{\text{A}})/w(s')}\nonumber \\
 & = & \frac{\mathcal{P}^{(w)}(s|I_{\text{A}})/\int_{\mathcal{S}}ds\,w(s'')\mathcal{\,P}(s''|I_{\text{A}})}{\int_{\mathcal{S}}ds\,\mathcal{P}^{(w)}(s'|I_{\text{A}})/\int_{\mathcal{S}}ds\,w(s''')\mathcal{\,P}(s'''|I_{\text{A}})}\nonumber \\
 & = & \mathcal{P}(s|I_{\text{A}}).
\end{eqnarray}
Here we first turned the attention back into a probability by inversely
weighting and re-normalization, then substituted $\mathcal{A}^{(w)}(s|I_{\text{B}})$
by $\mathcal{A}^{(w)}(s|I_{\text{A}})$ thanks to their identity,
further substituted the latter by its definition in terms of $\mathcal{P}(s|I_{\text{A}})$,
and finally used the normalization $\int_{\mathcal{S}}ds\,\mathcal{P}(s|I_{\text{A}})=1$.

The relative attention entropy differs from weighted relative entropy
(Eq.\ \ref{eq:weighted-relative-entropy}) due to the re-normalization
in the definition of attention, which leads to an irrelevant re-scaling
of weighted relative entropy, since independent of $I_{\text{B}}$,
but also to a relevant extra term that depends on $I_{\text{B}}$:
\begin{eqnarray}
\mathcal{D}_{s}^{(w)}(I_{\text{A}},I_{\text{B}}) & \negmedspace\!\negmedspace\!\negmedspace\!\negmedspace\!=\negmedspace\!\negmedspace\!\negmedspace\!\negmedspace\! & \frac{\mathcal{\widetilde{D}}_{s}^{(w)}(I_{\text{A}},I_{\text{B}})-\!\ln\frac{\int_{\mathcal{S}}\!ds\,w(s)\,\mathcal{P}(s|I_{\text{A}})}{\int_{\mathcal{S}}\!ds\,w(s)\,\mathcal{P}(s|I_{\text{B}})}}{\int_{\mathcal{S}}\!ds\,w(s)\,\mathcal{P}(s|I_{\text{A}})}\!\nonumber \\
\label{eq:relative-attention-entropy-2}
\end{eqnarray}
This extra term ensures properness when the attention entropy gets
minimized w.r.t.\ $I_{\text{B}}$. We refrained here to give similar
integration variables different names.

We investigate the specific scenario of Alice wanting to inform Bob
optimally. From this, we will motivate attention according to Eq.\ \ref{eq:relative-attention-entropy}.
This means that she wants to prepare Bob such that he can decide about
an action $a\in\mathbb{A}$ in a for him optimal way. This action
has an outcome for Bob that depends on the unknown situation $s$
Alice tries to inform Bob about. The outcome is described by a utility
function $u(s,a)$, which both, Alice and Bob want to maximize. The
utility depends on both, the unknown situation $s$ and Bob's action
$a$. In the following, we assume $u(s,a)$ to be at least twice differentiable
w.r.t.\ $a$ and to exhibit only one maxima for any given $s.$ In
case Alice does not know Bob's utility function, but its curvature
w.r.t.\ $a$ at its maxima for any given $s$, it will turn out that
Alice wants to communicate her attention $\mathcal{A}^{(w)}(s|I_{\text{A}})$
as accurately as possible to Bob, where $w(s)$ is the curvature of
the utility w.r.t.\ $a$. In short, we will show that Alice should
inform Bob most precisely about situations in which Bob's decision
requires the largest accuracy and not at all about situations in which
his actions do not make any difference. The functional, according
to which Alice will fit $I_{\text{B}}$ to $I_{\text{A}}$, will not
be the relative attention entropy of Eq.\ \ref{eq:relative-attention-entropy},
but a different one. However, it motivates attention as an essential
element of utility aware communication and shows the path to extend
the derivation of relative entropy to that of relative attention entropy.

\subsection{Attention Example\label{subsec:Attention-Example}}

\begin{figure*}[!t]
\includegraphics[width=1\textwidth]{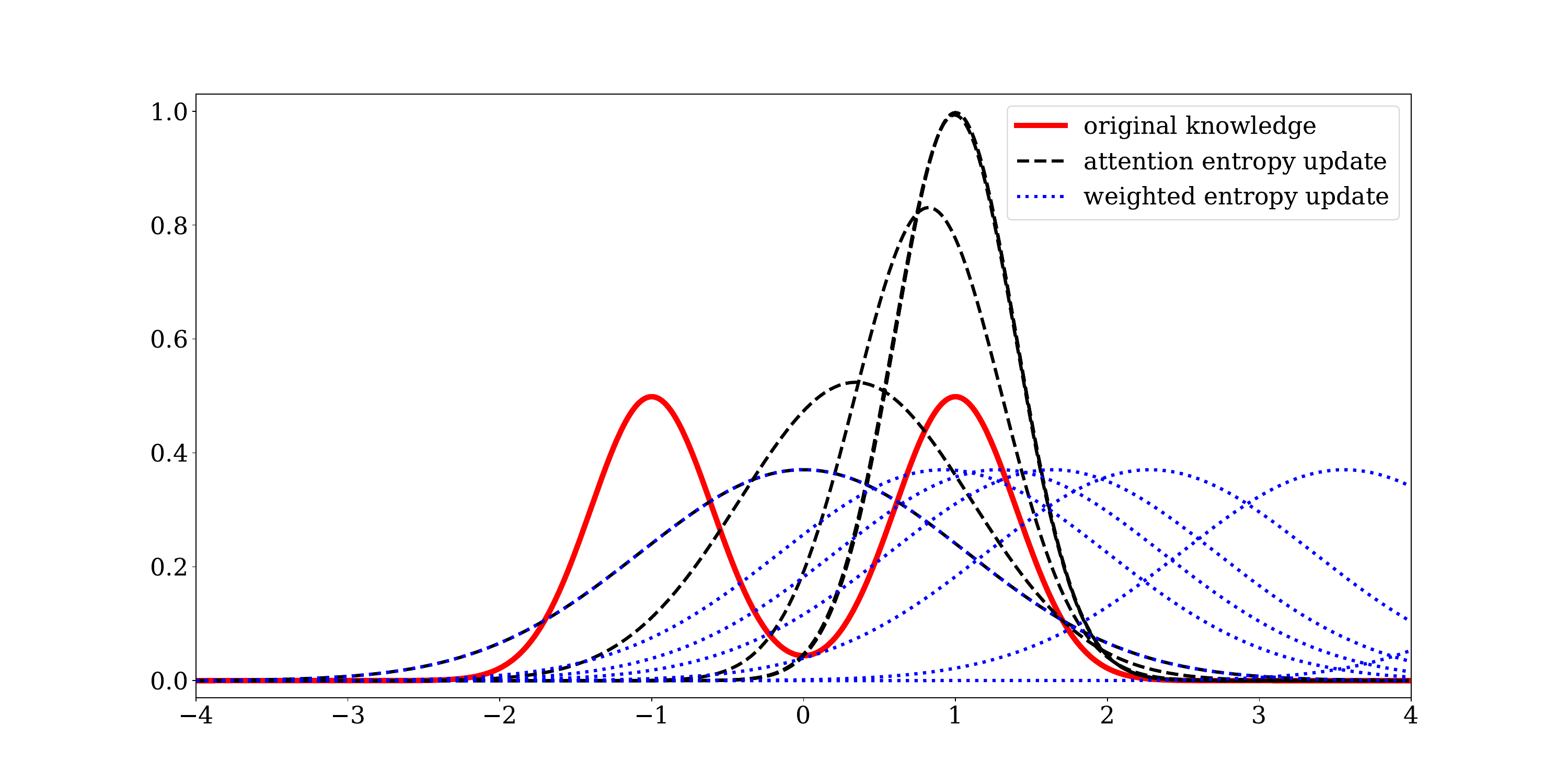}\caption{Example of communication based on relative attention entropy and weighted
relative entropy as discussed in Sect.\ \ref{subsec:Attention-Example}.
Alice's bimodal knowledge state is given by the red solid curve. Bob's
final knowledge state after Alice's communication is shown for various
cases. The dashed black lines correspond to cases in which Alice uses
relative attention entropy, and the dotted blue lines to cases she
uses weighted relative entropy. Different results for the weight function
$w(s)=\exp(\lambda\,s)$ with $\lambda=0$, $1$, $2$, $4,$ $8$,
$16$, and $32$ are shown from left two right, respectively. In case
$\lambda=0$ relative entropy, relative attention entropy, and weighted
relative entropy give the same result, the shown zero centered Gaussian.
For $\lambda\ge8$ the different curves for the relative attention
entropy results are visually indistinguishable and indicate the result
of the $\lambda\rightarrow\infty$ limit. \label{fig:Example-of-communication}}
\end{figure*}

\begin{figure*}
\includegraphics[viewport=100bp 50bp 1340bp 650bp,clip,width=0.5\textwidth]{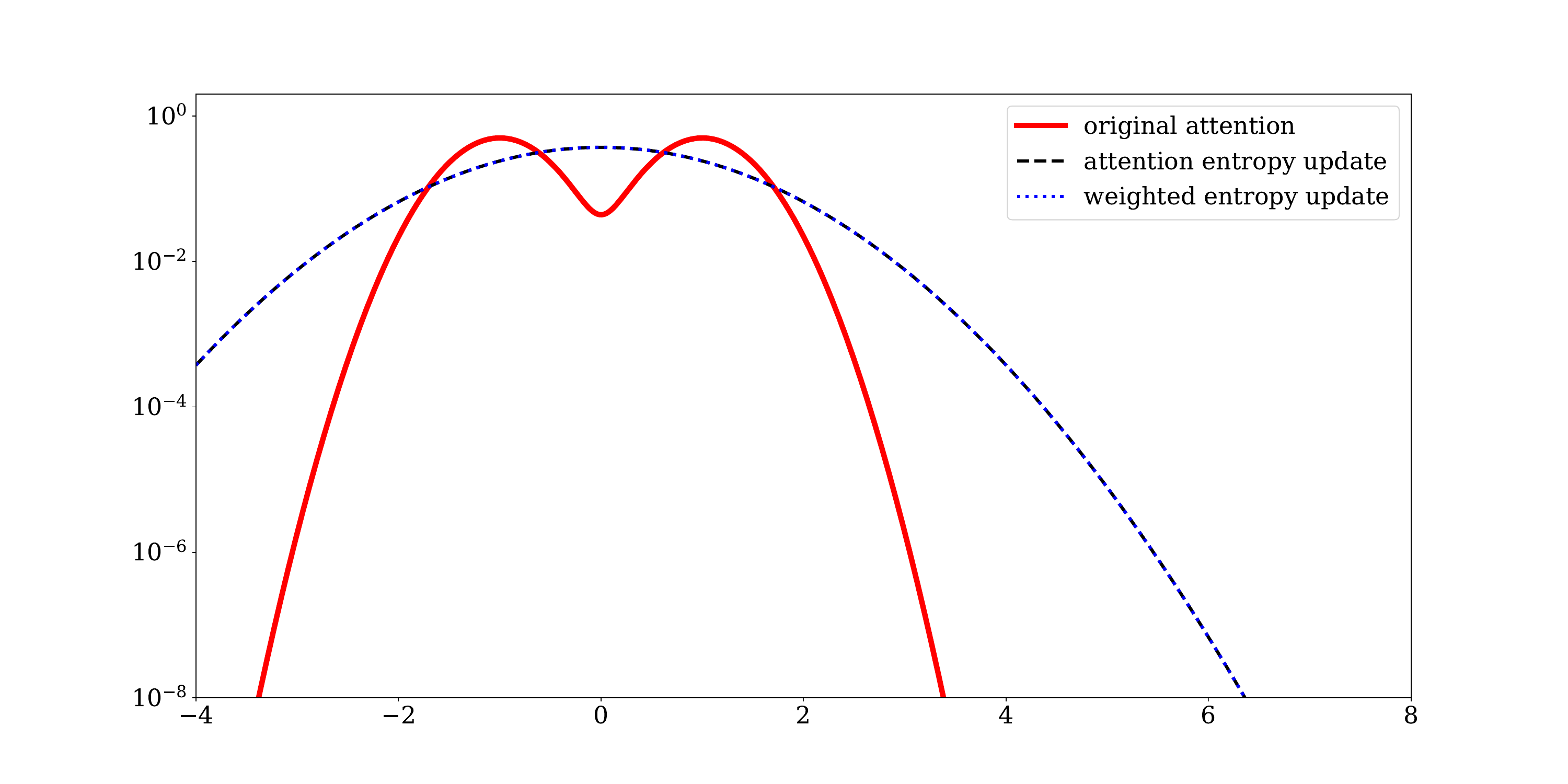}\includegraphics[viewport=100bp 50bp 1340bp 650bp,clip,width=0.5\textwidth]{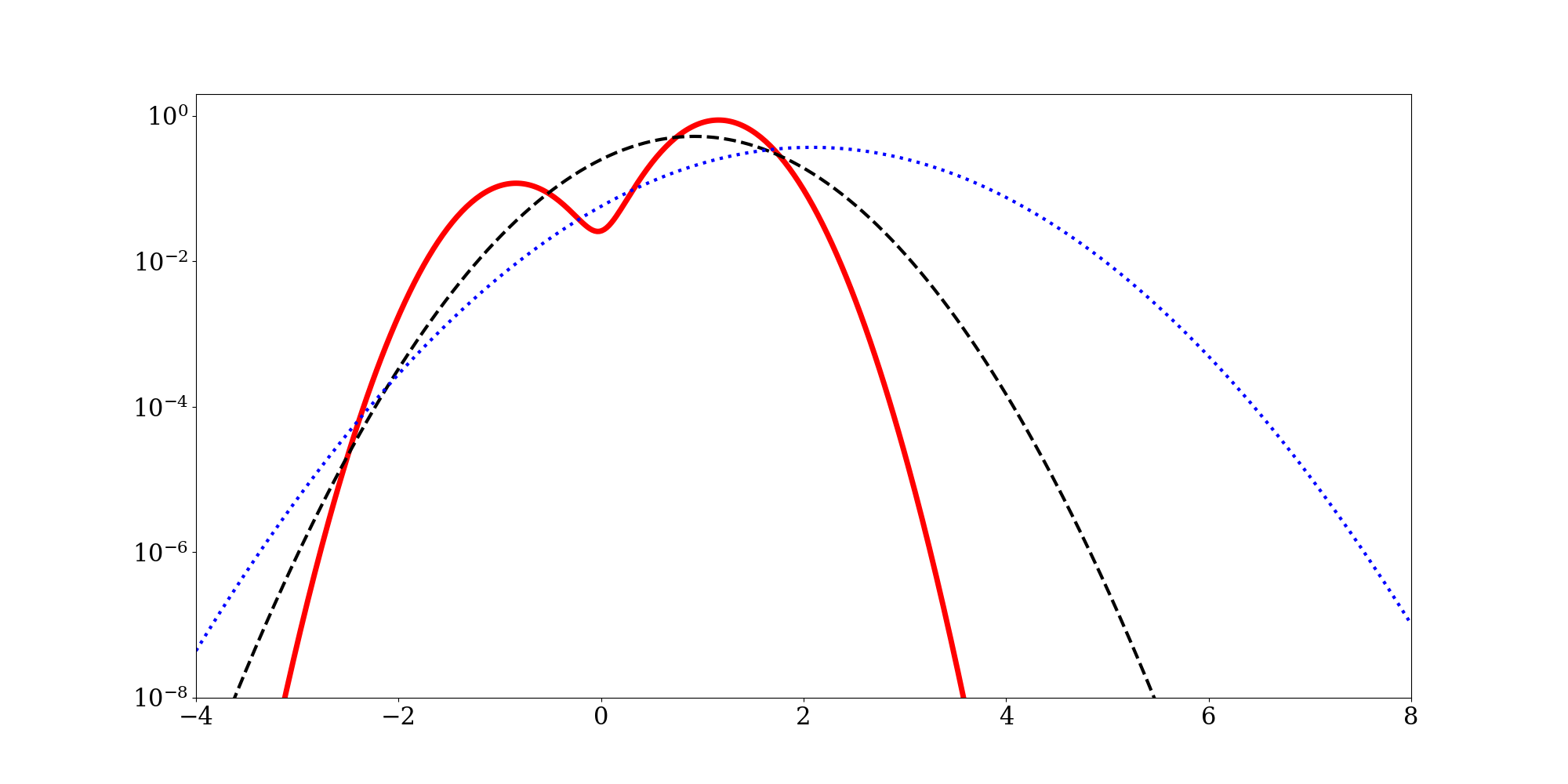}

\includegraphics[viewport=100bp 50bp 1340bp 650bp,clip,width=0.5\textwidth]{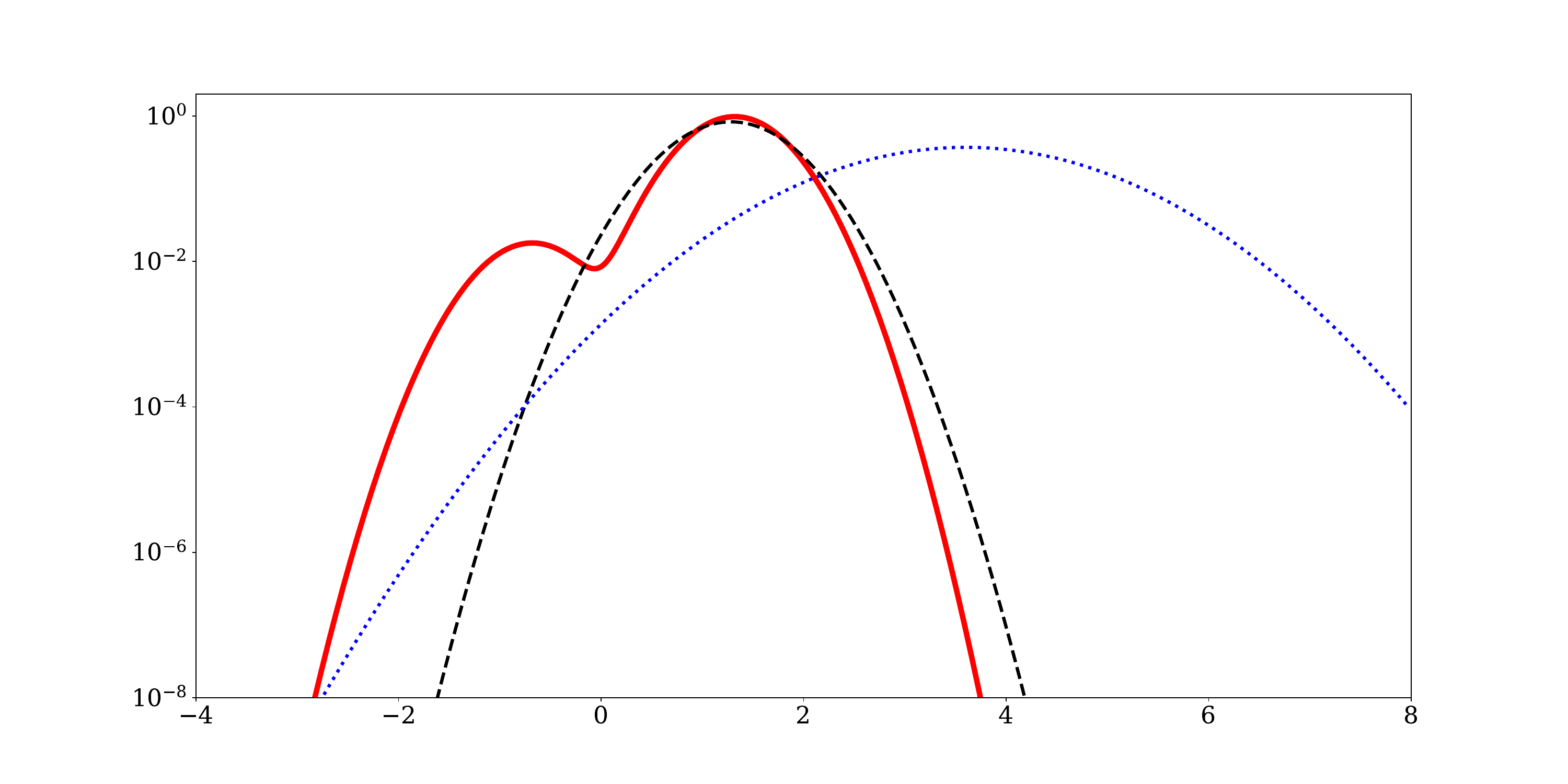}\includegraphics[viewport=100bp 50bp 1340bp 650bp,clip,width=0.5\textwidth]{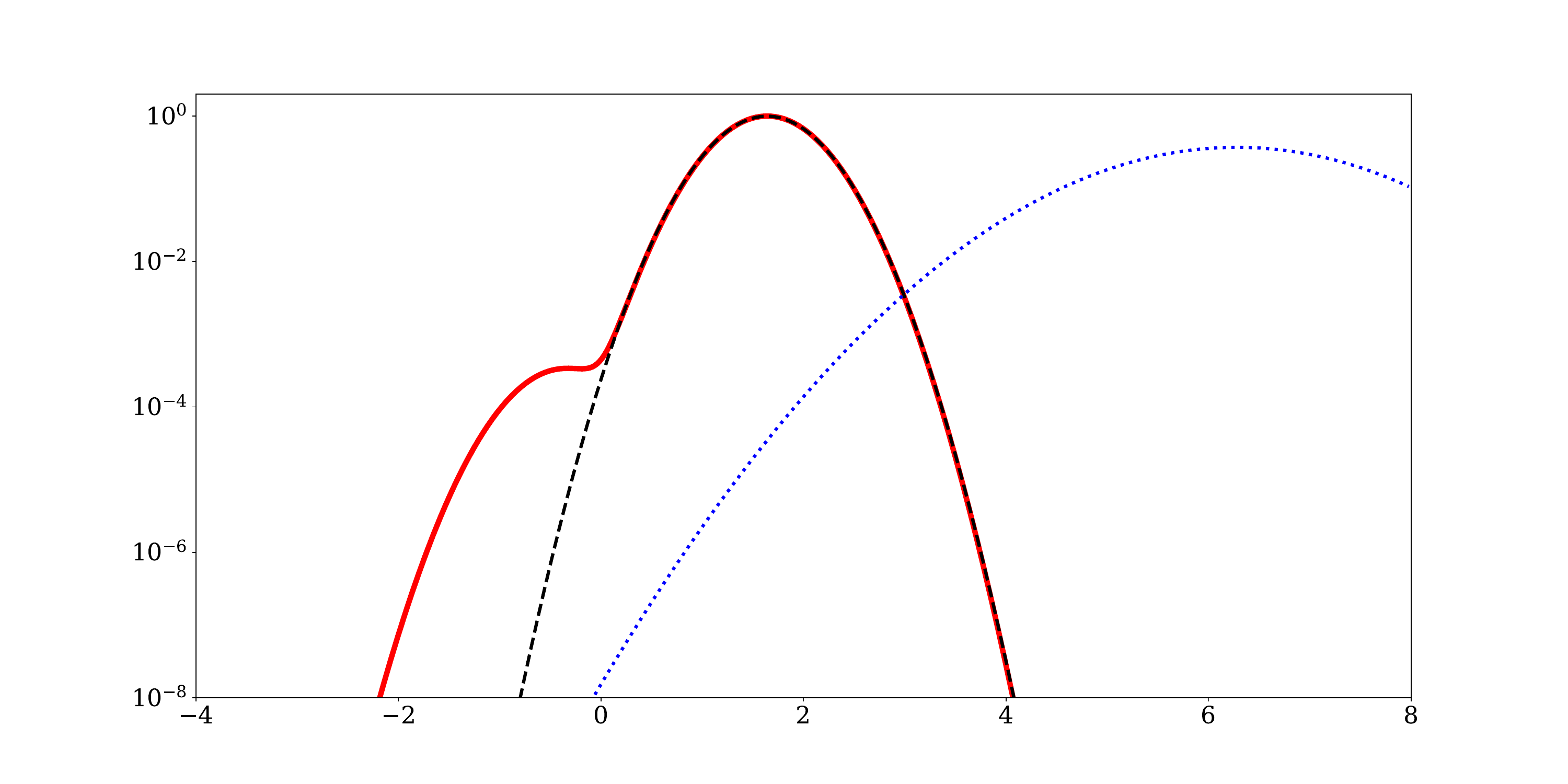}\caption{Attention functions corresponding to the cases $\lambda=0,\text{ }1,\text{ }2,\text{ and }4$
of Fig.\ \ref{fig:Example-of-communication} on logarithmic scale
to display the unattended peak of the Alice's attention. Note that
due to the strong exponential focus of the weights on larger $s$-values
the attention peaks are displaced to the right w.r.t.\ the corresponding
knowledge peaks. \label{fig:Attention-functions-corresponding}}
\end{figure*}

To illustrate how relative attention entropy works in practice, we
investigate an illustrative example. For this we assume that Alice
has a bimodal knowledge state 
\begin{equation}
\mathcal{P}(s|I_{\text{A}})=\frac{1}{2}\sum_{x\in\{-1,1\}}\mathcal{G}\left(s-x,\sigma_{\text{A}}^{2}\right)
\end{equation}
on a situation $s$ about which she wants to inform Bob, which is
a superposition of two Gaussians, with 
\begin{equation}
\mathcal{G}(s,\sigma^{2}):=\frac{1}{\sqrt{2\pi\sigma^{2}}}\exp\left(-\frac{s^{2}}{2\sigma^{2}}\right).
\end{equation}
This is shown in Fig.\ \ref{fig:Example-of-communication}.

Let us assume that Alice believes that the different situations have
importance weights $w(s)=e^{\lambda\,s}$ for Bob, with $\lambda$
controlling the inhomogeneity of the weights. We further assume that
her communication can only create Gaussian knowledge states in Bob
of the form
\begin{equation}
\mathcal{P}(s|I_{\text{B}})=\mathcal{G}\left(s-m,\sigma_{\text{B}}^{2}\right).
\end{equation}
The corresponding attentions of Alice and Bob are 
\begin{eqnarray}
\mathcal{A}^{(w)}(s|I_{\text{A}}) & = & \!\negmedspace\!\!\sum_{x\in\{-1,1\}}\negmedspace\!\negmedspace\!\frac{e^{x\,\lambda}\mathcal{G}\left(s-x-\lambda\sigma_{\text{A}}^{2},\sigma_{\text{A}}^{2}\right)}{2\,\cosh\lambda},\label{eq:A_A}\\
\mathcal{A}^{(w)}(s|I_{\text{B}}) & = & \mathcal{G}\left(s-m-\lambda\sigma_{\text{B}}^{2},\sigma_{\text{B}}^{2}\right),\label{eq:A_B}
\end{eqnarray}
respectively, as shown in App.\ \ref{sec:Attention-Example-Calculations},
which also covers the details of the following calculations. These
functions are displayed in Fig.\ \ref{fig:Attention-functions-corresponding}
for smaller values of $\lambda$.

Matching the parameters $m$ and $\sigma_{\text{B}}^{2}$ by minimizing
the relative attention entropy in Eq.\ \ref{eq:relative-attention-entropy}
w.r.t.\  those yield
\begin{eqnarray}
m & = & \tanh\lambda-\frac{\lambda}{\cosh^{2}\lambda},\label{eq:m}\\
\sigma_{\text{B}}^{2} & = & \sigma_{\text{A}}^{2}+1-\tanh^{2}\lambda.\label{eq:sigma_B}
\end{eqnarray}
The resulting communicated attention weighted knowledge is depicted
in Fig.\ \ref{fig:Example-of-communication} for various values of
$\lambda$. The case $\lambda=0$ corresponds to homogeneous weights
and therefore to using the non-weighted relative entropy, Eq.\ \ref{eq:relative-entropy}.
In this case Alice communicates a broad knowledge state to Bob that
covers both peaks of her knowledge. With increasing $\lambda$ the
right peak becomes more important and Alice puts more and more attention
on communicating its location and shape more accurately. In the limit
of $\lambda\rightarrow\infty$ we have $m=1$ and $\sigma_{\text{B}}^{2}=\sigma_{\text{A}}^{2}$,
meaning that Alice communicates exactly this relevant peak at $s=+1$
and completely ignores the for Bob irrelevant one at $s=-1$.

Minimizing the weighted entropy of Eq.\ \ref{eq:weighted-relative-entropy}
instead, yields $\mathcal{P}(s|\widetilde{I_{B}})=\mathcal{G}(s-\widetilde{m},\widetilde{\sigma}_{\text{B}}^{2})$
with
\begin{eqnarray}
\widetilde{m} & = & \lambda\,\sigma_{\text{A}}^{2}+\tanh\lambda\label{eq:m_re}\\
\widetilde{\sigma}_{\text{B}}^{2} & = & \sigma_{\text{A}}^{2}+1,\label{eq:sigma_re}
\end{eqnarray}
which results in poorly adapted communicated knowledge states, as
also depicted in Fig.\ \ref{fig:Example-of-communication}. Weighted
relative entropy just moves the broad peak centered originally at
zero for both entropies in case of no weight ($\lambda=0$) to increasingly
more extreme locations, which for $\lambda\rightarrow\infty$ become
completely detached from the location of the relevant peak. This detachment
is clearly visible in Fig.\ \ref{fig:Example-of-communication} and
Fig.\ \ref{fig:Attention-functions-corresponding}.

In order to see how the process of Alice informing Bob works in detail,
we have to understand how he decodes messages, as this defines the
format of the messages Alice can send. The best way for him to incorporate
knowledge sent to him into his own beliefs is by using the MEP, which
we assume he does in the following.

\subsection{Maximum Entropy Principle}

The MEP was derived as a device to optimally decode messages and to
incorporate their information into preexisting knowledge \cite{PhysRev.106.620,PhysRev.108.171,jaynes1963information,jaynes1968prior,jaynes03,paris1992method}.
The MEP states that among all possible updated probability distributions
that are consistent with the message, the one with the largest entropy
should be taken. Requiring that this update should be local, reparametrization
invariant (w.r.t.\ the way the signal or situation $s\in\mathcal{S}$
is expressed), and separable (w.r.t. handling independent dimensions
of $s$) enforces the functional form (up to affine transformations)
of this entropy to be 
\begin{equation}
D_{s}(I_{\text{B}},I_{0})=\int_{\mathcal{S}}ds\,\mathcal{P}(s|I_{\text{B}})\,\ln\frac{\mathcal{P}(s|I_{\text{B}})}{\mathcal{P}(s|I_{0})}
\end{equation}
 with $I_{0}$ being Bob's knowledge before and $I_{\text{B}}$ after
the update.\footnote{Jaynes original works on the MEP used Shannon entropy, and therefore
a uniform $\mathcal{P}(s|I_{0})$ in a discrete setting \cite{PhysRev.106.620,PhysRev.108.171},
whereas in his later works he introduced the relative entropy in particular
for the case of continuous probability densities \cite{jaynes1963information,jaynes1968prior,jaynes03}.}

We assume that Alice's message takes the form
\begin{equation}
``\langle f(s)\rangle_{(s|I_{A})}=d".\label{eq:statement}
\end{equation}
We call the function $f:\mathcal{S}\rightarrow\mathbb{R}^{n}$ the
\textbf{topic} of the communication, as it expresses the specific
aspects of $s$ that Alice's message is about. \emph{E.g.}\ in case
Alice wants to inform Bob about the first moment of the first component
of $s\in\mathbb{R}^{n}$, the topic would be $f(s)=s_{1}$. As Alice
communicates it, the topic is known to both, Alice and Bob. Here,
$\langle f(s)\rangle_{(s|I_{A})}:=\int_{\mathcal{S}}ds\,f(s)\,\mathcal{P}(s|I_{\text{A}})$
is a compact notation for evaluating this topic as an expectation
value. The communicated expectation value of $f$ is called in the
following the \textbf{message data} $d$. The word data expresses
that a quantity can be regarded as a certain fact, which still might
have an uncertain interpretation. In this case it is certain to Bob
that Alice claims that $\langle f(s)\rangle_{(s|I_{A})}$ has the
value $d$. Under the premise that Alice is honest, this informs Bob
about her belief, and if he can assume further that she is also well
informed, the data is informative on $s$ itself. Under the premise
that Alice is dishonest, $d$ could still inform Bob about what she
wants him to believe, and therefore still be data, just on Alice's
intentions and not on her knowledge.

Alice's message $m$ to Bob therefore consists of the tuple $(f,d).$
Although this message construction might look artificial at a first
glance, it actually can express many, if not all real world communications,
as we motivate in App.\ \ref{sec:Real-World-Communication}.

Updating his knowledge according to the MEP after receiving the message
implies that Bob extremizes the constrained entropy
\begin{eqnarray}
\!\!\!\!\!\mathcal{E}_{s}(I_{\text{B}},\mu|I_{0},m) & \!\!\!\!\!:=\!\!\!\!\! & \mathcal{D}_{s}(I_{\text{B}},I_{0})+\mu^{\text{t}}\left(\langle f(s)\rangle_{(s|I_{B})}-d\right)
\end{eqnarray}
w.r.t.\ the arguments $I_{\text{B}}$ and $\mu$. Here, the latter
is a Lagrange multiplier that ensures that Alice's statement, Eq.\ \ref{eq:statement},
is imprinted onto Bob's knowledge $\mathcal{P}(s|I_{\text{B}})$.
The MEP and the requirement of normalized probabilities then imply
that Bob's updated knowledge is of the form
\begin{eqnarray}
\mathcal{P}(s|I_{\text{B}}) & = & \frac{\mathcal{P}(s|I_{0})\,\exp\left(\mu^{\text{t}}f(s)\right)}{\mathcal{Z}(\mu,f)}\text{, with}\\
\mathcal{Z}(\mu,f) & := & \int\text{d}s\,\mathcal{P}(s|I_{0})\,\exp\left(\mu^{\text{t}}f(s)\right).
\end{eqnarray}
The Lagrange multiplier $\mu$ needs to be chosen such that $\langle f(s)\rangle_{(s|I_{\text{B}})}=d$,
which can be achieved by requiring
\begin{equation}
\frac{\partial\ln\mathcal{Z}(\mu,f)}{\partial\mu}=d,
\end{equation}
since 
\begin{eqnarray}
\frac{\partial\ln\mathcal{Z}(\mu,f)}{\partial\mu} & = & \frac{1}{\mathcal{Z}(\mu,f)}\frac{\partial\mathcal{Z}(\mu,f)}{\partial\mu}\nonumber \\
 & = & \frac{1}{\mathcal{Z}(\mu,f)}\int\text{d}s\,q(s)\,\exp\left(\mu^{\text{t}}f(s)\right)\nonumber \\
 & = & \langle f(s)\rangle_{(s|I_{\text{B}})}.
\end{eqnarray}
Thus, the MEP procedure ensures that the communicated moment of Alice's
knowledge state, $\langle f(s)\rangle_{(s|I_{\text{A}})}=d$, gets
transferred accurately into Bob's knowledge, 
\begin{equation}
\langle f(s)\rangle_{(s|I_{\text{A}})}=\langle f(s)\rangle_{(s|I_{\text{B}})},
\end{equation}
and thus that Bob extracts all information Alice has sent to him.
Alice's communicated expectation for the topic, $d=\langle f(s)\rangle_{(s|I_{\text{A}})}$,
is now ``imprinted'' onto Bob's knowledge, as $\langle f(s)\rangle_{(s|I_{\text{B}})}=d$
as well after his update.

If Bob decodes Alice's message via the MEP, she can send a perfectly
accurate image of her knowledge if the communication channel bandwidth
permits for this. Detailed explanations on this can be found in App.\ \ref{sec:Accurate-Communication}.
Otherwise, she needs to compromise and for this requires a criterion
on how to do so.

The protocol of the entropy-based communication between Alice and
Bob looks therefore as follows:

\paragraph*{Both know:}

The situation/signal $\ensuremath{s}$ is in a set $\ensuremath{\mathcal{S}}$
of possibilities, on which they have information $\ensuremath{I_{\text{A}}}$
and $\ensuremath{I_{0}}$, respectively, implying for each a knowledge
state $\mathcal{P}(s|I_{\text{A}})$ and $\mathcal{P}(s|I_{0})$,
respectively.

\paragraph*{Alice sends:}

A function $\ensuremath{f:\mathcal{S}\rightarrow\mathbb{R}^{n}}$,
the message topic, and her expectation for it, $\ensuremath{d=\langle f(s)\rangle_{P(s|I_{\text{A}})}}$,
the message data.

\paragraph*{Bob updates:}

$I_{0}\rightarrow I{}_{\text{B}}$, according to the MEP, such that
his updated knowledge $\mathcal{P}(s|I_{\text{B}})$ has the same
expectation value w.r.t. the topic, $\ensuremath{d=\langle f(s)\rangle_{P(s|I_{\text{B}})}}$.

\subsection{Structure of this Work}

The remainder of this work is structured as follows. Sect.\ \ref{sec:Proper-and-Weighted}
recapitulates the derivation of relative entropy by \textcite{Bernardo,2017Entrp..19..402L}
and shows that a non-trivially weighted relative entropy is not proper.
Sect.\ \ref{sec:Optimal-Communication} then discusses variants of
our communication scenario in which Alice's and Bob's utility functions
are known to them, be them aligned or misaligned. In Sect.\ \ref{sec:Attention}
we show that attention emerges as the quantity to be communicated
most accurately in case Alice wants the best for Bob, but lack's precise
knowledge on Bob's utility except for its curvature w.r.t\  Bob's
action in any situation. This motivates the introduction of attention
to entropic communication. In Sect.\ \ref{sec:Relative-Attention-Entropy}
relative attention entropy is derived in analogy to the derivation
of relative entropy. We conclude in Sect.\ \ref{sec:Conclusion}
with discussing the relevance of our analysis in technological and
socio-psychological contexts.

\section{Proper and Weighted Coding\label{sec:Proper-and-Weighted}}

\subsection{Proper Coding\label{subsec:Proper-Coding}}

In order to see the relation of being proper and relative entropy,
we recapitulate its derivation as given by \textcite{2017Entrp..19..402L}
in a modified way.\footnote{Note also that we modified the notation.}
There, it was postulated that Alice uses a loss function $\mathcal{L}(s,I_{\text{A}},I_{\text{B}})$
(negative utility) that depends on the situation $s$ that happens
in the end, as well as her and Bob's knowledge at this point, $I_{\text{A}}$
and $I_{\text{B}}$, respectively. \textcite{2017Entrp..19..402L}
call this function her embarrassment, as it should encode how badly
she informed Bob about the situation $s$ that happens in the end.
Obviously, Alice wants to minimize this embarrassment.

At the time Alice has to make her choice, she does not know which
situation $s=s_{*}$ will happen. She therefore needs to minimize
her expected loss
\begin{eqnarray}
\mathcal{L}(I_{\text{A}},I_{\text{B}}) & := & \left\langle \mathcal{L}(s_{*},I_{\text{A}},I_{\text{B}})\right\rangle _{(s_{*}|I_{\text{A}})}\label{eq:LL}
\end{eqnarray}
for deciding which knowledge state Bob $I_{\text{B}}$ should get
(via her message $m$). Here we discriminate the related, but different
functions labeled by $\mathcal{L}$ via their signatures (their sets
of arguments).

General criteria such a loss function should obey were formulated
by \cite{2017Entrp..19..402L}, which we slightly rephrase here as:
\begin{description}
\item [{Analytical:}] Alice's loss should be an analytic expression of
its arguments. An analytic function is an infinitely differentiable
function such that it has a converging Taylor series in a neighborhood
of any point of its domain. As a consequence, an analytical function
is fully determined for all locations of its domain (assuming this
is connected) by such a Taylor series around any of the domain positions.
\item [{Locality:}] In the end, only the case that happens matters. Without
loss of generality, let's assume that $s=s_{*}$ turned out to be
the case. Of all statements about $s$, only her prediction $\mathcal{P}(s_{*}|I_{\text{B}})$
that Alice made about $s_{*}$ before $s_{*}$ turned out to be the
case, is relevant for her loss.
\item [{Properness:}] If possible, Alice should favor to transmit her actual
knowledge state to Bob, $I_{\text{B}}=I_{\text{A}}$.
\item [{Calibration:}] The expected loss of being proper shall be zero.
\end{description}
\begin{enumerate}
\item \textbf{Locality} implies that $\mathcal{L}(s,I_{\text{A}},I_{\text{B}})$
can only depend on $I_{\text{A}}$ and $I_{\text{B}}$ through $q(s):=\mathcal{P}(s|I_{\text{B}})$
and $p(s):=\mathcal{P}(s|I_{\text{A}})$, meaning
\begin{equation}
\mathcal{L}(s,I_{\text{A}},I_{\text{B}})=\mathcal{L}(s,p(s),q(s))+\lambda\,\left(1-\int\text{d}s'\,q(s')\right),
\end{equation}
again using the function signatures\footnote{For $s\in\mathcal{S}$ given, $\mathcal{L}(s,I_{\text{A}},I_{\text{B}})$
depends on the full functions $p$ and $q$, whereas $\mathcal{L}(s,p(s),q(s))$
only depends on the values $p(s)$ and $q(s)$ of these functions
at that specific $s$.} to discriminate different $\mathcal{L}'$s and introducing a Lagrange
multiplier to ensure that $\mathcal{P}(s|I_{\text{B}})$ is normalized.
\end{enumerate}
\textbf{Properness} then requests that the expected loss should be
minimal for $q=p$, implying for all possible $s=s_{*}$

\begin{eqnarray}
0 & = & \left.\frac{\partial\mathcal{L}(I_{\text{A}},I_{\text{B}})}{\partial q(s_{*})}\right|_{q=p}\nonumber \\
 & = & \frac{\partial}{\partial q(s_{*})}\times\nonumber \\
 &  & \left.\left\langle \mathcal{L}(s,p(s),q(s))-\lambda\,\int\text{d}s'\,q(s')\right\rangle _{(s|I_{\text{A}})}\right|_{q=p}\nonumber \\
 & = & \left.\int\text{d}s\,\delta(s-s_{*})\left[\frac{\partial\mathcal{L}(s,p(s),y)}{\partial y}\right]_{y=q(s)}\!\!\!\!\mathcal{P}(s|I_{\text{A}})\right|_{q=p}\!\!\!\!-\lambda\nonumber \\
 & = & \left.\left[\frac{\partial\mathcal{L}(s,p(s_{*}),y)}{\partial y}\right]_{y=q(s_{*})}\!\!\!\!p(s_{*})\right|_{q=p}\!\!\!\!-\lambda\nonumber \\
 & = & \left.\left[\frac{\partial\mathcal{L}(s,x,y)}{\partial y}\right]_{y=x}x-\lambda\right|_{x=q(s_{*})}.\label{eq:properness}
\end{eqnarray}
From this 
\begin{equation}
\left[\frac{\partial\mathcal{L}(s,x,y)}{\partial y}\right]_{y=x}=\frac{\lambda}{x}\label{eq:condition}
\end{equation}
 follows, which is solved analytically by
\begin{equation}
\mathcal{L}(s,x,y)=\lambda\,\ln\,y+c(s,x),\label{eq:consequence}
\end{equation}
as can be verified by insertion. The Lagrange multiplier is unspecified
and we choose it to be $\lambda=1$, with the positive sign of $\lambda$
ensuring that $\mathcal{L}$ is actually a minimum and the magnitude
$|\lambda|=1$ that the units of this loss are nits ($\lambda=1/\ln2$
would set the units to bits or shannons).

\textbf{Calibration} requests then that $0=\mathcal{L}(s,x,x)=\ln\,x+c(s,x)$,
therefore $c(s,x)=-\ln x$, and thus, by reinsertion this into Eq.\ \ref{eq:LL},
we find
\begin{equation}
\mathcal{L}(I_{\text{A}},I_{\text{B}})=\int_{\mathcal{S}}ds\,\mathcal{P}(s|I_{\text{A}})\,\ln\frac{\mathcal{P}(s|I_{\text{A}})}{\mathcal{P}(s|I_{\text{B}})}.
\end{equation}
This is the relative entropy $\mathcal{D}_{s}(I_{\text{A}},I_{\text{B}})$
as defined by Eq.\ \ref{eq:relative-entropy}. We note that calibration
is more of an aesthetically requirement, as Alice's choice is already
uniquely determined by any loss functions that is local and proper.
Calibration, however, makes the loss reparametrization invariant,
as for any diffeomorphism $s'=s'(s)$ we find $\mathcal{D}_{s'}(I_{\text{A}},I_{\text{B}})=\mathcal{D}_{s}(I_{\text{A}},I_{\text{B}})$,
as can be verified by a coordinate transformation:

\begin{eqnarray}
\mathcal{D}_{s'}(I_{\text{A}},I_{\text{B}}) & \!\!\!\!=\!\!\!\! & \int_{\mathcal{S}}ds'\,\mathcal{P}(s'|I_{\text{A}})\,\ln\frac{\mathcal{P}(s'|I_{\text{A}})}{\mathcal{P}(s'|I_{\text{B}})}\nonumber \\
 & \!\!\!\!=\!\!\!\! & \int_{\mathcal{S}}ds\,\underbrace{\left|\frac{ds'(s)}{ds}\right|\mathcal{P}(s'(s)|I_{\text{A}})}_{=\mathcal{P}(s|I_{\text{A}})}\,\ln\frac{\mathcal{P}(s'(s)|I_{\text{A}})\left|\frac{ds}{ds'}\right|}{\mathcal{P}(s'(s)|I_{\text{B}})\left|\frac{ds}{ds'}\right|}\nonumber \\
 & \!\!\!\!=\!\!\!\! & \int_{\mathcal{S}}ds\,\mathcal{P}(s|I_{\text{A}})\,\ln\frac{\mathcal{P}(s|I_{\text{A}})}{\mathcal{P}(s|I_{\text{B}})}\nonumber \\
 & \!\!\!\!=\!\!\!\! & \mathcal{D}_{s}(I_{\text{A}},I_{\text{B}})
\end{eqnarray}

Strictly speaking, Eq.\ \ref{eq:condition} implies Eq.\ \ref{eq:consequence}
only for an infinitesimal environment of $y=x$. Only thanks to the
requirement of the loss being \textbf{analytic} in the full domain
of its second argument (added here to the requirement set of \textcite{2017Entrp..19..402L}),
Eq.\ \ref{eq:consequence} has to hold at all other locations and
Alice's expected loss becomes uniquely determined to be the relative
entropy.

\subsection{Weighted Coding\label{subsec:Weighted-Coding}}

Could one insert weights into the relative entropy by inserting those
into the above derivation? One could try to do so by modifying the
locality requirement by requiring
\begin{eqnarray}
\mathcal{L}^{(w)}(s,I_{\text{A}},I_{\text{B}}) & \!\!\!\!:=\!\!\!\! & w(s)\,\mathcal{L}^{(w)}(s,p(s),q(s))\nonumber \\
 &  & +\lambda\,\left(1-\int\text{d}s'\,q(s)\right)
\end{eqnarray}
to have a minimal expectation value for Alice. Propriety requires
that
\begin{equation}
\left[\frac{\partial\mathcal{L}^{(w)}(s,x,y)}{\partial y}\right]_{y=x}=\frac{\lambda}{x\,w(s)},
\end{equation}
as follows from a calculation along the lines of Eq.\ \ref{eq:properness}.
This is analytically solved by
\begin{eqnarray}
\mathcal{L}^{(w)}(s,x,y) & = & \frac{1}{w(s)}\,\ln\,\frac{x}{y},
\end{eqnarray}
where we directly ensured calibration and set $\lambda=1$. Alice's
expected weighted loss is therefore (for normalized $q(s)=\mathcal{P}(s|I_{\text{B}})$)
\begin{eqnarray}
\mathcal{L}^{(w)}(I_{\text{A}},I_{\text{B}})\!\!\!\! & \!\!\!\!\!\!\!\!\!=\!\!\!\!\!\!\!\!\!\! & \!\!\!\!\left\langle \mathcal{L}^{(w)}(s,I_{\text{A}},I_{\text{B}})\right\rangle _{(s|I_{\text{A}})}\!\!\!\!\!\!\!\!+\lambda\,\overbrace{\left(\int\text{d}s'\,q(s)-1\right)}^{=0}\nonumber \\
\!\!\!\! & = & \!\!\!\!\left\langle w(s)\,\mathcal{L}^{(w)}(s,\mathcal{P}(s|I_{\text{A}}),\mathcal{P}(s|I_{\text{B}}))\right\rangle _{(s|I_{\text{A}})}\nonumber \\
\!\!\!\! & = & \!\!\!\!-\left\langle \,\frac{w(s)}{w(s)}\,\ln\,\frac{\mathcal{P}(s|I_{\text{A}})}{\mathcal{P}(s|I_{\text{B}})}\right\rangle _{(s|I_{\text{A}})}\nonumber \\
\!\!\!\! & = & \!\!\!\!\mathcal{L}(I_{\text{A}},I_{\text{B}}),
\end{eqnarray}
which is exactly the same unweighted relative entropy $\mathcal{D}_{s}(I_{\text{A}},I_{\text{B}})$
as before. Therefore, weights in a relative entropy as a means of
deciding on optimal coding are not consistent with our requirements.
Since we have modified the locality requirement, and since calibration
is not essential, the requirement that prevents weights must be properness.

As weighted relative entropy is not proper \cite{gkelsinis2020theoretical},
we ask whether a different way to introduce weights could be proper.
In order to answer this question, we turn to a conceptually simpler
setting. For this we introduce the concept of Theory of Mind \cite{premack1978does,baron1985does}.
This is the representation of a different mind in one's own mind.
As this can be applied recursively (``\emph{I think that you think
that I think ...}'') one discriminates Theory of Minds of different
orders according to the level of recursion. The above derivation of
the relative entropy is based on a second order Theory of Mind construction,
namely that Alice does not want Bob to think badly about her informedness
(\emph{She worries about} \emph{his beliefs on her thinking}). A first
order Theory of Mind construction, in which Alice only cares about
Bob being well informed for what matters for Bob, might be more instructive
to understand how weights might emerge. We now turn to such a scenario.

\section{Optimal Communication\label{sec:Optimal-Communication}}

\subsection{Communication Scenario}

\begin{figure}[!t]
\includegraphics[clip,width=1\columnwidth]{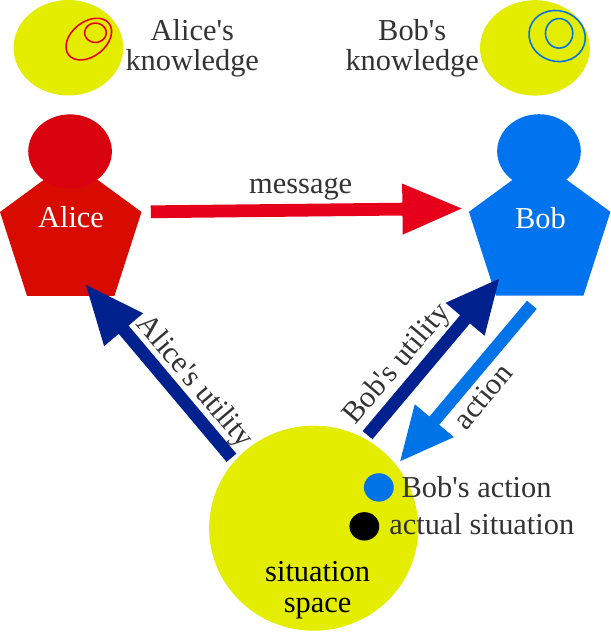}\caption{Sketch of the investigated communication scenario. Alice communicates
parts of her knowledge to Bob about an unknown situation. After updating
according to Alice's message Bob chooses his action, which for simplicity
is here assumed to be a point in situation space $\mathcal{S}$ (for
example, the blue point could indicate to which situation his action
is best adapted). His action and the unknown situation determine Bob's
resulting utility. Bob chooses his action by maximizing his expected
utility given his knowledge after Alice informing him (blue equal
probability contours of $\mathcal{P}(s|I_{\text{B}})$ in his mental
copy of the situation space). The action and situation also determine
a utility for Alice, which may or may not equal Bob's utility. Alice
chooses her message such that her expected utility resulting from
Bob's action is maximized in the light of her situation knowledge
$\mathcal{P}(s|I_{\text{A}})$ (red contours). In case she is honest,
she can only choose which parts of her knowledge she reveals with
her message by deciding on a message topic $f(s)$; the message data
is then determined to be $d=\langle f(s)\rangle_{(s|I_{\text{A}})}$.\label{fig:Sketch}}
\end{figure}

In the following we investigate the scenario sketched in Fig.\ \ref{fig:Sketch}:
Alice is relatively well informed about the situation $s\in\mathcal{S}$
her communication partner Bob will find himself in. In this not perfectly
known situation Bob needs to take an action $a\in\mathbb{A}$, which
is rewarded by a utility $u_{\text{B}}(s_{*},a)\in\mathbb{R}$ to
Bob that depends on the situation $s_{*}\in\mathcal{S}$ that will
eventually occur. Bob's action also implies a utility $u_{\text{A}}(s_{*},a)\in\mathbb{R}$
to Alice. We will eventually assume that Alice's and Bob's utility
functions are aligned, $u_{\text{A}}(s,a)=u_{\text{B}}(s,a)=:u(s,a)$,
and give arguments why this might happen, but for the moment, we keep
them separate in order to study the consequences of different interests.

Such misalignments are actually very common. Just imagine that Bob
is Alice's young child, $a$ is the amount of sugar that Bob is going
to consume, and $s$ is how well his metabolism handles sugar. There
is a lot of anecdotal evidence for $u_{\text{A}}(s,a)\neq u_{\text{B}}(s,a)$
under such conditions.

Alice will communicate through a bandwidth limited channel parts of
her knowledge to Bob, who is here assumed to trust Alice fully, so
that Bob can perform a more informed decision on his action. As stated
before, Alice's message takes the form $``\langle f(s)\rangle_{(s|I_{A})}=d"$,
with $f:\mathcal{S}\rightarrow\mathbb{R}^{n}$, the conversation topic,
being a moment function over the set of situations out of a limited
set $\mathcal{F}$ of such functions she can choose from.

With increasing set size of $\mathcal{F}$ Alice's communication channel
becomes more flexible and with increasing dimension $n$ of the data
space the channel bandwidth increases. Her message $m$ to Bob therefore
consists of the tuple $(f,d)\in\mathcal{F}\times\mathbb{R}^{n}$.\footnote{There is a bit of redundancy in this notation, since $\mathcal{U}(m):=(U(f),U(d))$
for any invertible affine $U:\mathbb{R}^{n}\rightarrow\mathbb{R}^{n}$
encodes the same information as $m=(f,d)$. For example, a modification
of the topic allows to absorb the information of the data $d$ into
the topic by using $U(f)=f-d$, which changes the message to $m'=\mathcal{U}(m)=(U\,(f),U\,(d))=(f-d,0)$.
As this redundancy is not a problem for the following discussion we
leave it in the notation.}

In case 
\begin{equation}
\langle f(s)\rangle_{(s|I_{A})}=d
\end{equation}
actually holds as communicated, Alice is honest, otherwise she lies.

Alice assumes Bob to perform a knowledge update $I_{0}\overset{m}{\rightarrow}I_{\text{B}}$
following the MEP upon receiving her message $m=(f,d)$ such that
\begin{eqnarray}
\langle f(s)\rangle_{(s|I_{\text{B}})} & = & d.
\end{eqnarray}
Thus, by choosing $f$ and $d$ Alice can directly determine certain
expectation values of Bob's knowledge, which then influence Bob's
action.

\subsection{Optimal Action\label{subsec:Optimal-Action}}

The action Bob chooses optimally is
\begin{eqnarray}
a_{\text{B}} & = & \underset{a\in\mathbb{A}}{\text{argmax}}\,u_{\text{B}}(a|I_{\text{B}})\text{, with}\\
u_{B}(a|I_{\text{B}}) & = & \langle u_{\text{B}}(s,a)\rangle_{(s|I_{\text{B}})}.
\end{eqnarray}
For simplicity we assume that this action is uniquely determined and
that $u_{B}(a|I_{B})$ is twice differentiable w.r.t.\ $a$. Then,
we find that $a_{\text{B}}$ solves 
\begin{eqnarray}
0 & = & \left.\frac{\partial u_{\text{B}}(a|I_{\text{B}})}{\partial a}\right|_{a=a_{\text{B}}}=\left\langle g_{\text{B}}(s,a_{\text{B}})\right\rangle _{(s|I_{\text{B}})}\!\!\text{,}\label{eq:uB-optimal}\\
g_{\text{B}}(s,a) & := & \frac{\partial u_{\text{B}}(s,a)}{\partial a},
\end{eqnarray}
meaning that for the action $a_{\text{B}}$ Bob chooses, his expectation
for his utility gradient $g_{\text{B}}(s,a)$ w.r.t. his action has
to vanish.

From Alice's perspective, Bob's optimal action $a$ would be
\begin{eqnarray}
a_{\text{A}} & = & \underset{a\in\mathbb{A}}{\text{argmax}}\,u_{\text{A}}(a|I_{\text{A}})\text{, with}\label{eq:Alice_preference}\\
u_{\text{A}}(a|I_{\text{A}}) & = & \langle u_{\text{A}}(s,a)\rangle_{(s|I_{\text{A}})},
\end{eqnarray}
implying that $a_{\text{A}}$ solves 
\begin{eqnarray}
0 & = & \left.\frac{\partial u_{\text{A}}(a|I_{\text{A}})}{\partial a}\right|_{a=a_{\text{A}}}=\left\langle g_{\text{A}}(s,a_{\text{A}})\right\rangle _{(s|I_{\text{A}})}\!\!,\label{eq:Alice_preference2}\\
g_{\text{A}}(s,a) & := & \frac{\partial u_{\text{A}}(s,a)}{\partial a}.
\end{eqnarray}

\subsection{Dishonest Communication}

Let's first investigate the scenario of Alice being so eager to manipulate
Bob's action to her advantage that she is willing to lie for that.
In order that Bob does what Alice finds optimal for herself, $a_{\text{B}}=a_{\text{A}}$,
she needs to manipulate Bob's updated knowledge state $I_{\text{B}}$
such that 
\begin{equation}
\left\langle g_{\text{B}}(s,a_{\text{A}})\right\rangle _{(s|I_{\text{B}})}=0,\label{eq:moment-transfer}
\end{equation}
according to Eq.\ \ref{eq:uB-optimal}. Thus, it would be advantageously
for her if Bob's expected utility gradient $g_{\text{B}}(s,a_{\text{A}})$
vanishes for the action $a_{\text{A}}$, which Alice prefers him to
take, since then he would take this action, $a_{\text{B}}=a_{\text{A}}$.
Alice can achieve this by setting his expectation for $g_{\text{B}}(s,a_{\text{A}})$
to zero via communicating him an appropriate deceptive message. Any
message $m=(f,d)$ with the topic being $f(s)=g_{\text{B}}(s,a_{\text{A}})+c$
and the data $d=c$ will achieve that Eq.\ \ref{eq:moment-transfer}
is satisfied. Here $c\in\mathbb{R}^{n}$ is an arbitrary constant
Alice might use to ensure $f\in\mathcal{F}$ (if this is possible)
or to obscure her manipulation. As her deceptive communication derives
from Alice's knowledge and utility, these quantities imprint onto
her message. This happens through the usage of her optimal action
$a_{\text{A}}$ in $g_{\text{B}}(s,a)$ in Eq.\ \ref{eq:moment-transfer}.
This by her preferred action is specified by Eq.\ \ref{eq:Alice_preference}.

Thus, Alice will use a communication topic $f$ that reflects Bob's
interest, as it is built on $g_{\text{B}}(\cdot,a)$, however evaluated
for Alice's preferred action $a_{\text{A}}$ in this scenario. In
order for her manipulation to work, she has to make Bob believe that
$\left\langle g_{\text{B}}(s,a_{\text{A}})\right\rangle _{(s|I_{\text{B}})}=0$
and hope that this will let him indeed choose $a=a_{\text{A}}$.

Interestingly, Alice does not need to know Bob's initial knowledge
state $I_{0}$ for this, as the MEP update ensures that the relevant
moment of Bob's updated knowledge $I_{\text{B}}$ gets the necessary
value, see Eq.\ \ref{eq:moment-transfer}. Nevertheless, she needs
to know his interests $g_{\text{B}}(s,a)$, as through exploiting
those she can manipulate his action.

In the likely case of $d\neq\left\langle f(s)\right\rangle _{(s|I_{\text{A}})}$,
Alice would be lying. However, lying is risky for Alice, since Bob
might detect her lies in the long run, being it for Bob's knowledge
after Alice informing him $\mathcal{P}(s|I_{\text{B}})$ turns out
too often to be a bad predictor for $s$, or by other telltale signs
of Alice. Bob realizing that Alice lies could have negative consequences
for her in the long run, therefore we assume in the following that
Alice is always honest. However, she might still follow her own interests.\footnote{In case Bob realizes Alice is lying, he will stop updating his knowledge
according to her messages, and she will loose her ability to inform
him w.r.t.\ $s$, for good or bad. For the complex dynamics that
can emerge under not fully honest communication, the reader is referred
to \cite{2022AnP...53400277E}. Bob might even decode from her message
partly what Alice's interests are, as $a_{\text{A}}$ and $g_{\text{A}}(s,a_{\text{A}})=\partial u_{\text{A}}(s,a_{\text{A}})/\partial a_{\text{A}}$
imprint onto her message's topic. This can even enable him to choose
actions that deviate from Alice's interests as largely as he can afford
in order to punish her for lying. Thus, although lying can definitively
bring a short term advantage to Alice, in the long run it could cost
Alice more than she might gain in the beginning. For this reason,
she might decide to become honest or to be honest already in the first
place. Although performing punishments typically costs Bob in terms
of his own utility, he might choose them in a way that they cost Alice
more than himself. This way, they might educate her to become honest,
which would then let them be a good investment for Bob, as he will
benefit from the information Alice can share.}

\subsection{Topics under Misaligned Interests\label{subsec:Misaligned-Interests}}

What Alice faces in the general case of differing interests is a complex
mathematical problem. Even if Alice is bound to be honest, she still
has some influence on Bob by deciding which part of her knowledge
she shares. She does not need to give him information that would drive
his decision against her own interests. By choosing the conversation
topic smartly, Alice could make Bob acting in a way that is beneficial
to both of them to some degree.

Let us assume for now that Alice knows both utility functions, Bob's
and her own, as well as Bob's initial knowledge $\mathcal{P}(s|I_{0})$.
For a given $f\in\mathcal{F}$ used as the topic of her honest communication
$m=(f,\langle f(s)\rangle_{(s|I_{\text{A}})})$, she can work out
Bob's resulting updated knowledge $I_{\text{B}}$=$I_{\text{B}}(f)$,
his action $a$, as well as how advantageous that action would be
for her, by calculating and optimizing
\begin{eqnarray}
u_{\text{A}}(f) & := & \left\langle u_{\text{A}}(s,a_{\text{B}}(f))\right\rangle _{(s|I_{\text{A}})}\text{, with}\label{eq:uA(f)}\\
a_{\text{B}}(f) & := & \underset{a\in\mathbb{A}}{\text{argmax}}(u_{\text{B}}(a|I_{\text{B}}(f))),\label{eq:aB}\\
u_{\text{B}}(a|I_{\text{B}}) & := & \left\langle u_{\text{B}}(s,a)\right\rangle _{(s|I_{\text{B}})},\\
\mathcal{P}(s|I_{\text{B}}(f)) & = & \frac{\mathcal{P}(s|I_{0})\,\exp\left(\mu\,f(s)\right)}{\mathcal{Z}(\mu,f)},\label{eq:P(s|I_B(f))}\\
\mathcal{Z}(\mu,f) & = & \int\text{d}s\,\mathcal{P}(s|I_{0})\,\exp\left(\mu\,f(s)\right)\text{, with}\\
\mu=\mu(f):\,d & \!\!\!\!\!\!\!\!\!\!\!\!=\!\!\!\!\!\!\!\!\!\!\!\! & \frac{\partial\ln\mathcal{Z}(\mu,f)}{\partial\mu}.\label{eq:mu}
\end{eqnarray}

The last step here, determining $\mathcal{P}(s|I_{\text{B}})$, is
also an optimization problem according to the MEP. Thus, the optimal
topic for Alice therefore results from the three fold nested optimization

\begin{eqnarray}
f & \!\!=\!\! & \underset{f}{\text{argmax}\,}u_{\text{A}}(\underset{a}{\text{argmax}}\,u_{\text{B}}(a|f,\underset{I_{\text{B}}:\langle f(s)\rangle_{I_{\text{B}}}=\langle f(s)\rangle_{I_{\text{A}}}\!\!\!\!\!\!\!\!\!\!\!\!\!\!\!\!\!\!\!\!\!\!\!\!\!\!\!\!\!\!\!}{\text{argmax}}(\mathcal{D}_{s}(I_{\text{B}},I_{0}))).\nonumber \\
\end{eqnarray}

Analytic solutions to this can only be expected in special cases.
For future reference and numerical approaches to the problem, we calculate
the relevant gradient $\partial u_{\text{A}}/\partial f$ in App.\ \ref{sec:Topic-Gradient}.
Its component for $s_{*}$ is
\begin{eqnarray}
\frac{\text{d}u_{\text{A}}(f)}{\text{d}f(s_{*})}\!\!\!\! & = & \!\!\!\!-\left\langle g_{\text{A}}(s,a)\right\rangle _{(s|I_{\text{A}})}^{\text{t}}\left[\frac{\partial^{2}u_{\text{B}}(a|I_{\text{B}})}{\partial a\partial a^{\text{t}}}\right]^{-1}\times\label{eq:topic_gradient}\\
 &  & \!\!\!\!\left\{ g_{\text{B}}(s_{*},a)\,\mathcal{P}(s_{*}|I_{\text{B}})\,\mu^{\text{t}}-\left\langle g_{\text{B}}(s,a)\delta f(s)^{\text{t}}\right\rangle _{(s|I_{\text{B}})}\right.\nonumber \\
 &  & \!\!\!\!\left.\times\left\langle \delta f(s)\,\delta f(s)^{\text{t}}\right\rangle _{(s|I_{\text{B}})}^{-1}\left[\mathcal{P}(s_{*}|I_{\text{B}})-\mathcal{P}(s_{*}|I_{\text{A}})\right]\right\} ,\nonumber 
\end{eqnarray}
where $a=a_{\text{B }}(f)$, $\mathcal{P}(s_{*}|I_{\text{B}})=\mathcal{P}(s_{*}|I_{\text{B}}(f))$,
and $\mu=\mu(f)$ are given by Eq.\ \ref{eq:aB}, \ref{eq:P(s|I_B(f))},
and \ref{eq:mu}, respectively, and $\delta f(s):=f(s)-\left\langle f(s)\right\rangle _{(s|I_{\text{B}})}$.
Inspection of the condition $\partial u_{\text{A}}/\partial f=0$
and the terms that could allow for it is instructive, as it hints
to the factors that drive Alice's topic choice. Alice has found a
local optimal topic for her communication when either
\begin{itemize}
\item $g_{\text{A}}(a|I_{\text{A}}):=\left\langle g_{\text{A}}(s,a)\right\rangle _{(s|I_{\text{A}})}=0$,
meaning that Alice's remaining interest $g_{\text{A}}(a|I_{\text{A}})$
is perfectly satisfied by Bob's resulting action $a$,
\item for any situation $s_{*}$ a sophisticated balance $b(s_{*}):=\left\{ \text{terms in curly brackets of Eq.\ \ref{eq:topic_gradient}}\right\} =0$
holds between Bob's interest $g_{\text{B}}(s_{*},a)$ in that situation
and the difference in the probabilities Alice and Bob assign to it,
$\mathcal{P}(s_{*}|I_{\text{B}})-\mathcal{P}(s_{*}|I_{\text{A}})$,
or
\item the not balanced term $b(s_{*})$ is orthogonal to Alice's remaining
interest $g_{\text{A}}(a|I_{\text{A}})$ w.r.t.\ a metric given by
the inverse Hessian of Bob's expected utility $u_{\text{B}}(a|I_{\text{B}})$
(as derived w.r.t.\ his action $a$),
\end{itemize}
and the corresponding location is not a minimum of the expected utility.

Detailed investigation of the general case of misaligned interests
are left for future work. Here, only an illustrative example will
be examined.

\subsection{Example of Misaligned Interests \label{subsec:Example-of-Misaligned}}

\begin{figure*}[!t]
\includegraphics[viewport=0bp 50bp 1440bp 680bp,clip,width=1\textwidth]{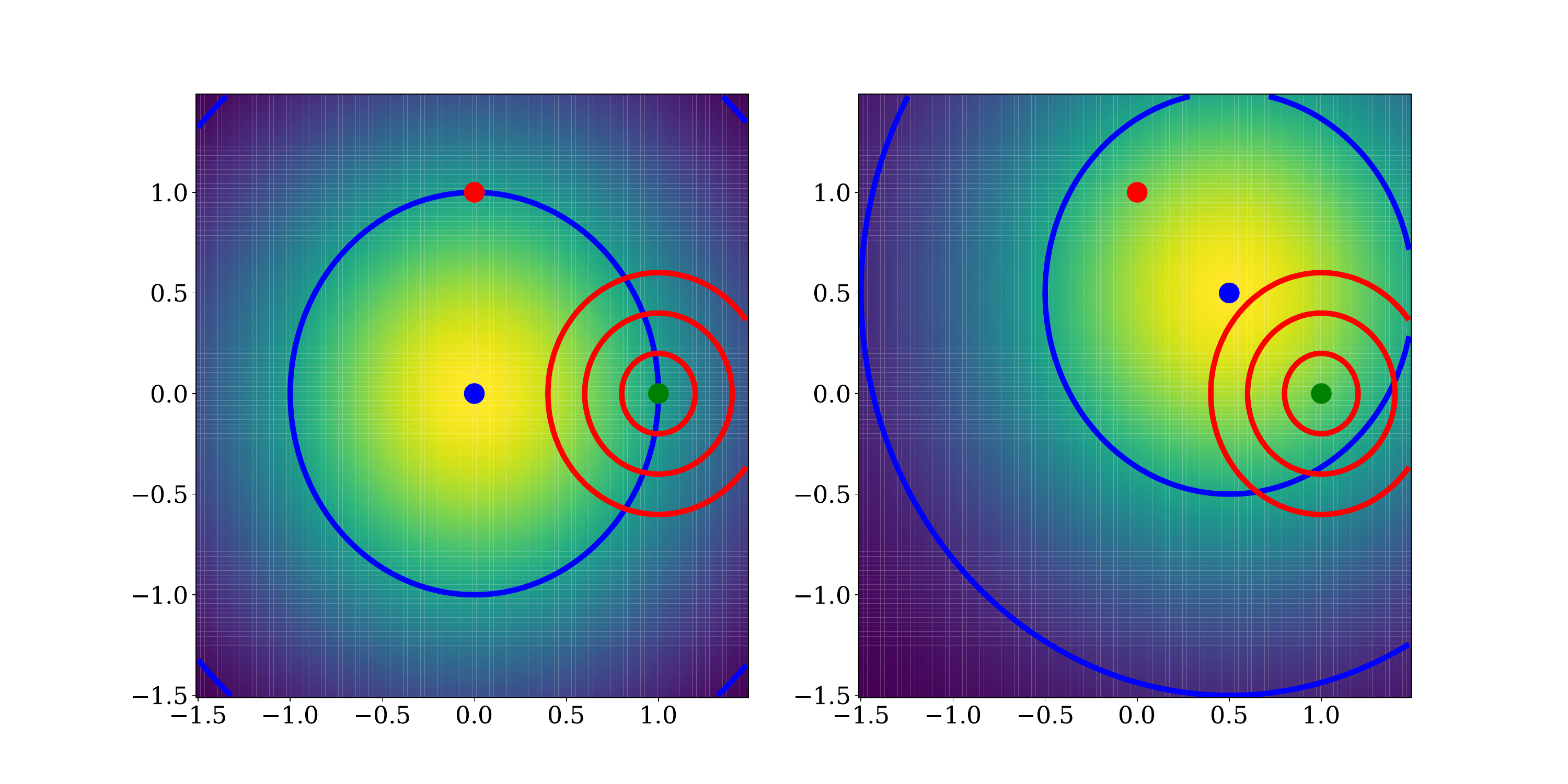}

\caption{Knowledge states and preferred actions of Alice and Bob in case of
misaligned interests before (left) and after (right) the communication.
The plane of $s$-values is shown. Bob's knowledge state initially,
$\mathcal{P}(s|I_{0})$ (left), and finally, $\mathcal{P}(s|I_{\text{B}})$
(right), is shown by the background color as well as by the blue contour
lines at the 1- and 2-sigma levels. Alice's more precise knowledge
is indicated only via red 1-, 2-, and 3-sigma level contours. The
dots mark possible actions for Bob that are optimal for him under
his knowledge (blue), under Alice's knowledge (green), or optimal
for Alice (red). Comparing the two panels, especially the movement
of Bob's optimal action (blue dot) between them, shows that Alice
informs Bob such that he chooses an action that is a compromise between
their interests.\label{fig:Knowledge-states-and}}
\end{figure*}

An instructive example of misaligned interests is in order. For this,
let us assume that the space of possible situations as well as that
of actions have two dimensions, $s,a\in\mathbb{R}^{2}$. Alice's and
Bob's initial beliefs shall be Gaussian distributions 
\begin{eqnarray}
\mathcal{P}(s|I'_{\text{A/B}}) & = & \mathcal{G}(s-\overline{s}_{\text{A/B}},S_{\text{A/B}})\text{, with}\\
\mathcal{G}(s,S) & := & \frac{1}{||2\pi S||^{\nicefrac{1}{2}}}\exp\left(-\frac{1}{2}s^{\text{t}}S^{-1}s\right),
\end{eqnarray}
$\overline{s}_{\text{A}}=(1,0)^{\text{t}}$, $\overline{s}_{\text{B}}=(0,0)^{\text{t}}$,
and $S_{\text{A}}<S_{\text{B}}=\mathbb{1}$ (spectrally), so that
Alice is better informed than Bob (since $S_{\text{A}}<S_{\text{B}})$
and the chosen coordinate system is aligned with that knowledge ($s_{1}$-axis
is parallel to $\overline{s}_{\text{A}}$). Furthermore, Bob's utility
should be 
\begin{equation}
u_{\text{B}}(s,a)=-\left\Vert s-a\right\Vert ^{2},
\end{equation}
so that he wants his action to match the situation. Alice would prefer
if he matched a\emph{ }by an angle $\varphi$ rotated target according
to her utility
\begin{eqnarray}
u_{\text{A}}(s,a) & = & -\left\Vert R\,s-a\right\Vert ^{2}\text{, with}\\
R & = & \begin{pmatrix}\cos\varphi\,\, & -\sin\varphi\\
\sin\varphi\, & \cos\varphi
\end{pmatrix}
\end{eqnarray}
establishing a misalignment of their interests.

In this situation, Bob's expected utility is 
\begin{equation}
u_{\text{B}}(a|I_{\text{B}})=-\left\langle s^{\text{t}}s\right\rangle _{(s|I_{\text{B}})}+2a^{\text{t}}\left\langle s\right\rangle _{(s|I_{\text{B}})}-a^{\text{t}}a,
\end{equation}
from which his action
\begin{equation}
a_{\text{B}}=\left\langle s\right\rangle _{(s|I_{\text{B}})}
\end{equation}
follows. Thus, the first moment of Bob's belief on $s$ determines
fully his action and therefore Alice only needs to inform him about
that. Let us therefore assume that Alice will use a topic of the form
$f(s)=\tau^{\text{t}}\,s$, with $\tau=(\cos\alpha,\sin\alpha)^{\text{t}}$
some normalized direction, so that $\tau^{\text{t}}\tau=1$. We check
later using Eq.\ \ref{eq:topic_gradient} whether this is her best
choice or not.

The data of her message is then
\begin{equation}
d=\tau^{\text{t}}\left\langle s\right\rangle _{(s|I_{\text{A}})}=\tau^{\text{t}}\overline{s}_{\text{A}}=\cos\alpha
\end{equation}
and Bob's updated knowledge state becomes
\begin{eqnarray}
\mathcal{P}(s|I_{\text{B}}) & = & \mathcal{G}(s-d\,\tau,\mathbb{1}),
\end{eqnarray}
as verified in the following: The MEP fixes Bob's updated knowledge
to be of the form 
\begin{eqnarray}
\mathcal{P}(s|I_{\text{B}}) & = & \frac{\exp\left(-\frac{1}{2}s^{\text{t}}s+\mu\,\tau^{\text{t}}s\right)}{\text{\ensuremath{2\ensuremath{\pi}}}\mathcal{Z}}\text{ with}\\
\mathcal{Z} & = & e^{\mu^{2}/2}.
\end{eqnarray}
Requiring that the communicated moment is matched leads to
\begin{eqnarray}
\ln\mathcal{Z} & = & \frac{\mu^{2}\,}{2},\\
\frac{\partial\ln\mathcal{Z}}{\partial\mu} & = & \mu=d=\tau^{\text{t}}\overline{s}_{\text{A}}=\cos\alpha\text{, and}\\
\mathcal{P}(s|I_{\text{B}}) & = & \frac{\exp\left(-\frac{1}{2}\left[s^{\text{t}}s-2\,d\,\tau^{\text{t}}s+d^{2}\right]\right)}{\text{2\ensuremath{\pi}}}\nonumber \\
 & = & \mathcal{G}(s-d\,\tau,\mathbb{1}),
\end{eqnarray}
as we claimed. This implies $a_{\text{B}}=d\,\tau=\tau\,\tau^{\text{t}}\overline{s}_{\text{A}}$.
Thus in this situation Bob does exactly what Alice tells him, as $a_{\text{B}}=d\,\tau$
consists of the two essential elements of her message $m=(\tau^{\text{t}}s,d)$.
However, being honest, Alice is not fully free in what she can say.
With choosing the topic direction $\tau$ the message's data $d=\tau^{\text{t}}\overline{s}_{\text{A}}$
are fully determined thanks to her honesty.

Therefore, Alice's expected utility 
\begin{eqnarray}
u_{\text{A}}(a) & = & -\left\langle s^{\text{t}}R^{\text{t}}Rs\right\rangle _{(s|I_{\text{A}})}+2d^{\text{t}}R\overline{s}_{\text{A}}-d^{\text{t}}d\nonumber \\
 & = & -\text{\text{t}r}\left\langle ss^{\text{t}}\right\rangle _{(s|I_{\text{A}})}+2\overline{s}_{\text{A}}^{\text{t}}\tau\,\tau^{\text{t}}\,R\overline{s}_{\text{A}}-\overline{s}_{\text{A}}^{\text{t}}\overline{s}_{\text{A}}\nonumber \\
 & = & -\text{\text{t}r}(S_{\text{A}}+\overline{s}_{\text{A}}\overline{s}_{\text{A}}^{\text{t}})-\overline{s}_{\text{A}}^{\text{t}}\overline{s}_{\text{A}}+2\overline{s}_{\text{A}}^{\text{t}}\tau\,\tau^{\text{t}}\,R\overline{s}_{\text{A}}\nonumber \\
 & = & -\text{\text{t}r}(S_{\text{A}})+2\cos\alpha\,\left(\cos\alpha\cos\varphi+\sin\alpha\sin\varphi\right)\nonumber \\
\end{eqnarray}
becomes maximal for the topic direction

\begin{eqnarray}
\alpha & = & \arctan\left[\frac{-\cos\varphi\pm1}{\sin\varphi}\right],
\end{eqnarray}
as a straightforward calculation shows. The sign of $\pm$ has to
be chosen such that $u_{\text{A}}(a)$ is maximal, which for $\varphi\in[-\pi,\pi]$
turns out to be $+$.

For mostly aligned interests, $\varphi\ll1$, Alice's optimal topic
has an angle of $\alpha=\frac{1}{2}\varphi+\mathcal{O}(\varphi^{3})$,
which means it is a nearly perfect compromise between what is optimal
for her and for Bob. For $\alpha=0$ their interests are perfectly
aligned and Alice informs Bob ideally with the statement ``$\langle s_{1}\rangle_{(s|I_{\text{A}})}=1$''.

In case of orthogonal interests, $\varphi=\pi/2$, Alice's optimal
topic angle is $\alpha=\pi/4$, informing effectively with a statement
like ``$\langle s_{1}+s_{2}\rangle_{(s|I_{\text{A}})}=1$'', which,
given that $\langle s\rangle_{(s|I_{\text{A}})}=(1,0)^{\text{t}}$,
is less informative for Bob than the statement ``$\langle s_{1}\rangle_{(s|I_{\text{A}})}=1$''
Alice would have made under aligned interests. Bob's resulting decision
of $a=(\nicefrac{1}{2},\nicefrac{1}{2})^{\text{t}}$ in that case
turns out to be a perfect compromise between their interests, or put
differently being sub-optimal for each of them to the same degree.
This situation is depicted in Fig.\ \ref{fig:Knowledge-states-and}.

For anti-aligned interests, $\varphi=\pi$, Alice's optimal topic
angle is $\alpha=\pi/2$ as then she can only send the uninformative
message ``$\langle s_{2}\rangle_{(s|I_{\text{A}})}=0$'', as revealing
any more of her knowledge would be against her interests. This leaves
Bob's knowledge unchanged and therefore lets him pick the action $a=\overline{s}_{\text{B}}=(0,0)^{\text{t}}$.

To summarize, misalignment of interests leads to a communication and
a resulting action that are a compromise between the interests of
the communication partners. Who of the two has to compromise more
depends on details of their knowledge states and their interests.

Alice informing Bob sub-optimally to her own advantage bears for Alice
the risk of Bob realizing this. In repeated situations, Bob might
recognize the systematic misalignment angle $\alpha=\text{angle}(s,a)$
between the $s_{*}$ that happened and the topic direction $\tau$
chosen by Alice on the basis of her information on $s$. This might
either let him question Alice's good intentions for him or her competence.
In either case he could threaten Alice to ignore or even counteract
against her advice until he gets convinced that she has largely aligned
her interests with his.

The general optimal topic of Alice's message could, however, be a
non-linear function of the situation instead of the linear $f(s)=\tau^{\text{t}}s$
assumed above. This can be checked by inspecting the functional gradient
of $u_{\text{A}}(f)$ w.r.t.\ $f$ as given by Eq.\ \ref{eq:topic_gradient}.
In case it vanishes for all $s\in\mathcal{S}$ the topic was optimal.

It turns out that this gradient only vanishes when Alice's and Bob's
interests are aligned ($\varphi=0$), as shown in App.\ \ref{sec:Specific-topic-gradient}.
For the instructive case of orthogonal interests ($\varphi=\nicefrac{\pi}{2}$),
however, the gradient 
\begin{eqnarray}
\frac{\text{d}u_{\text{A}}(f)}{\text{d}f(s)}\!\!\!\! & = & \!\!\!\!\frac{s_{2}-s_{1}}{\sqrt{2}}\,\mathcal{P}(s|I_{\text{B}})
\end{eqnarray}
does not vanish. This indicates that she could construct a more sophisticated
message that would pull Bob's resulting action a bit closer towards
her own interest $a_{\text{A}}$ and further away from the action
optimal for him (under her knowledge). The precise form of the optimal
topic for Alice is left for future work.

\subsection{Aligned Interests}

In the following we assume that Alice simply wants the best for Bob
from his perspective ($a_{\text{A}}=a_{\text{B}}$) and therefore
adapts his utility for herself, 
\[
u_{\text{A}}(s,a)=u_{\text{B}}(s,a)=:u(s,a).
\]

In this case, Alice informs Bob optimally via the message $m=(f,d)$
with $f(s)=g_{\text{A}}(s,a_{\text{A}})+c=g_{\text{B}}(s,a_{\text{B}})+c$,
which leads to a synchronization of their expectations w.r.t.\ the
most relevant moment of $s,$ $\langle f(s)\rangle_{(s|I_{\text{A}})}=\langle f(s)\rangle_{(s|I_{\text{B}})}$,
and therefore to an alignment of their optimal actions, $a_{\text{A}}=a_{\text{B}}=:a_{*}$.

For simplicity, we assume in the following that the action is described
by one real number $a\in\mathbb{R}$. An extension to a vector valued
action space $\mathcal{A}=\mathbb{R}^{u}$ is straightforward, but
does not add too much to the discussion below except of complexity
in the notation. Furthermore we assume Alice and Bob's common utility
function to be uni-modal in a given situation and to be well approximated
within the relevant region by
\begin{equation}
u(s,a)=v(s)-\frac{1}{2k}\left(\frac{a-b(s)}{\sigma(s)}\right)^{2k}.\label{eq:power-loss}
\end{equation}
Here, $b(s)$ is the optimal action in a given situation $s$, $v(s)$
the utility of this optimal action, $\sigma(s)\in\mathbb{R^{+}}$
the tolerance for deviation from the optimal action, and $k\in\mathbb{N}$
is specifying how harsh larger deviations reduce the utility. This
should serve as a sufficiently generic model that can capture a large
variety of realistic situations. In particular the case of a quadratic
loss (=negative utility) with $k=1$ mimics the typical situation
in which a Taylor expansion in $a$ around the optimal action $b(s)$
can be truncated after the quadratic term.

Alice's expected utility of Bob's action
\begin{equation}
u(a|I_{\text{A}})=\langle u(s,a)\rangle_{(s|I_{\text{A}})}
\end{equation}
has the gradient
\begin{eqnarray}
\frac{\partial u(a|I_{\text{A}})}{\partial a} & = & -\left\langle \frac{\left[a-b(s)\right]^{2k-1}}{\left[\sigma(s)\right]^{2k}}\right\rangle _{(s|I_{\text{A}})}\\
 & = & -\sum_{n=0}^{2k-1}b_{n}a^{n},\text{with}\\
b_{n} & = & \begin{pmatrix}2k-1\\
n
\end{pmatrix}\left\langle \frac{\left[-b(s)\right]^{2k-1}}{\left[\sigma(s)\right]^{2k}}\right\rangle _{(s|I_{\text{A}})}.
\end{eqnarray}
This is a polynomial of odd order in $a$ and therefore guaranteed
to have at least one real root. The maximum of $u(a|I_{\text{A}})$
among all such roots then gives the optimal action $a_{*}$. Thus,
the topic function
\[
f(s)=\frac{\left[a_{*}-b(s)\right]^{2k-1}}{\left[\sigma(s)\right]^{2k}}+c
\]
is Alice's best choice for a communication that ensures that Bob makes
an optimal decision.

In case $b(s)=s$ and $\sigma(s)=1$ this is $f(s)=(a_{*}-s)^{2k-1}$,
which is a polynomial of order $2k-1$ in $s$. Instead of communicating
the expectation value of this polynomial, which requires her to work
out $a_{*}$, Alice could simply communicate all moments up to order
$2k-1$ and thereby ensure that Bob would have all information needed
in order to decide on the optimal action.

Here, the requirement of properness appears in a weak form. Alice
wanting to inform Bob on a number of moments of her knowledge in order
to put him into a position to make a good decision is a weak form
of the requirement of properness. Full properness would be that Alice
wants to inform Bob to know all possible moments of her knowledge.
Thus, properness is expected to occur when Alice does not know Bob's
utility function, but wants to support him no matter what his interests
are. We will now turn to such a scenario.

\section{Attention\label{sec:Attention}}

We saw why Alice might align her interest with Bob's and in the following
assume this to have happened, $u_{\text{A}}=u_{\text{B}}\equiv u$.
Her knowledge on Bob's utility function influences how she selects
her message optimally. For the concept of attention to appear in her
reasoning, Alice must not know Bob's utility function in detail. In
case she did, she would optimize for the utility function. However,
she needs to be aware of the sensitivity with which Bob's utility
reacts to Bob's choices in the different situations $s$ in order
to give those situations appropriate weights in her communication.
These weights will determine how accurately she should inform about
the different situations such that Bob is optimally prepared to make
the right decision.

To be concrete, let Alice assume that in a given situation $s$ Bob's
utility has a single maximum at some to her unknown optimal action
$a_{*}(s)=b(s)$ of a to her unknown height $u(s,a_{*}(s))=v(s)$,
but with a to her known curvature $w(s)$. Furthermore, she assumes
that this utility function can be well Taylor-approximated around
any of these maxima as
\begin{equation}
u(s,a,v,w,b)=v(s)-\frac{1}{2}w(s)\left(a-b(s)\right)^{2}.\label{eq:power-loss-1}
\end{equation}
This corresponds to the case of Eq.\ \ref{eq:power-loss} with $k=1$
and $\sigma(s)=\left[w(s)\right]^{-1/2}$. We have added the parameters
$v$, $w$, and $b$ to the list of arguments of this approximate
utility function as Alice needs to average over the ones unknown to
her, which are $v$ and $b$.

In order to circumvent the technical difficulty to deal with probabilities
over functional space let us restrict the following discussion to
discrete $s\in\mathcal{S}.$ In this case the parameters $v$, $w$,
and $b$ become finite dimensional vectors with components $v_{s}=v(s)$,
etc.. The case of a continuous set of situations is dealt with in
App.\ \ref{sec:Attention-in-Continuous}.

Alice assumes that Bob's action will depend on these parameters and
is given by
\begin{eqnarray}
a_{*}(v,b,\sigma,I_{B}) & = & \text{argmax}_{a}\left\langle u(s,a,v,w,b)\right\rangle _{(s|I_{\text{B}})}\nonumber \\
 & = & a:\left[\left\langle \frac{\partial u(s,a,v,w,b)}{\partial a}\right\rangle _{(s|I_{\text{B}})}=0\right]\nonumber \\
 & = & a:\left[-\left\langle \left[a-b(s)\right]\,w(s)\right\rangle _{(s|I_{\text{B}})}=0\right]\nonumber \\
 & = & \frac{\left\langle b(s)\,w(s)\right\rangle _{(s|I_{\text{B}})}}{\left\langle w(s)\right\rangle _{(s|I_{\text{B}})}}\nonumber \\
 & =: & \left\langle b(s)\right\rangle _{(s|I_{\text{B}})}^{(w)},
\end{eqnarray}
where $a:\left[f(a)=0\right]$ means the (here assumed to be unique)
value of $a$ that fulfills $f(a)=0$. Furthermore, we introduced
the $w$-weighted expectation value
\begin{equation}
\left\langle b(s)\right\rangle _{(s|I_{\text{B}})}^{(w)}:=\sum_{s\in\mathcal{S}}\,b(s)\,\mathcal{A}^{(w)}(s|I_{\text{B}})
\end{equation}
that involves the attention $\mathcal{A}^{(w)}(s|I_{\text{B}})$ defined
in Eq.\ \ref{eq:attention}.

Let us assume that Alice believes that the curvature of Bob's utility
maxima are $w_{*}(s)$. This might be because she can estimate the
influence Bob's actions have on his own well being in the different
situations. For example, in the extreme case that Bob might be dead
in a given $s_{*}$, she might set $w_{*}(s_{*})=0$ as none of the
possible actions of Bob then matter any more to him. It will turn
out that the actual values of $w(s)$ do not matter, only their relative
values w.r.t.\ each other.

Thus, her knowledge about Bob's utility is 
\begin{equation}
\mathcal{P}(v,b,w|I_{\text{A}})=\delta(w-w_{*})\,\mathcal{P}(v,b|I_{\text{A}}),
\end{equation}
with $\mathcal{P}(v,b|I_{\text{A}})$ a relatively uninformative probability
density. We assume this to be independent on her knowledge on the
situation, $\mathcal{P}(s,v,b|I_{\text{A}})=\mathcal{P}(s|I_{\text{A}})\mathcal{\,P}(v,b|I_{\text{A}})$.

Furthermore, we assume here, in order to have a simple instructive
scenario, that Alice only has a vague idea around which value $\overline{b}:=\langle b(s)\rangle_{(b,s|I_{\text{A}})}$
the location of the maximum $b(s)$ of Bob's utility could be, and
how much it could deviate from $\overline{b}$. We assume that she
is not aware of any correlation of this function nor any structure
of its variance and therefore define
\begin{eqnarray}
\overline{b}(s) & := & \left\langle b(s)\right\rangle _{\!(b|I_{\text{A}})}=\overline{b},\\
D(s,s') & := & \left\langle \left(b(s)-\overline{b}(s)\right)\,\left(b(s')-\overline{b}(s)\right)\right\rangle _{\!(b|I_{\text{A}})}\\
 & =: & \gamma^{2}\,\delta_{s\,s'}.
\end{eqnarray}
In the last step we used that Alice's knowledge on $b$ is uninformative,
therefore unstructured, and thus its uncertainty covariance proportional
to the unit matrix. The parameter $\gamma$ expresses how much variance
Alice expects in $b$. Its precise value will turn out to be irrelevant.

Other setups in which $\overline{b}(s)$ is not a constant or $D(s,s')$
contain cross-correlations, are addressed in App.\ \ref{sec:Attention-in-Continuous}.

With the above assumptions, the expected utility is
\begin{eqnarray}
u(I_{\text{B}}) & = & \left\langle u\left(s,a_{*}(v,w,b),v,w,b\right)\right\rangle _{\!(s,v,w,b|I_{\text{A}})}\label{eq:utility-w-mariginalized}\\
 & = & \left\langle u\left(s,\left\langle b(s)\right\rangle _{(s|w,I_{\text{B}})},v,w,b\right)\right\rangle _{\!(s,v,b,\sigma|I_{\text{A}})}\nonumber \\
 & = & \left\langle v(s)-\frac{w(s)}{2}\left(\left\langle b(s)\right\rangle _{(s|w,I_{\text{B}})}-\overline{b}\right)^{2}\right\rangle _{\!\!(s,v,w,b|I_{\text{A}})}\nonumber \\
 & = & \underbrace{\left\langle v(s)\right\rangle _{(s,v|I_{\text{A}})}}_{=:V(I_{\text{A}})}-\frac{1}{2}\underbrace{\left\langle w_{*}(s)\right\rangle _{(s|I_{\text{A}})}}_{=:w_{I_{\text{A}}}}\times\nonumber \\
 &  & \left\langle \left(\left\langle b(s)\right\rangle _{(s|w_{*},I_{\text{B}})}-\overline{b}\right)^{2}\right\rangle _{(s,b|I_{\text{A}})}^{(w_{*})}\nonumber \\
 & = & V(I_{\text{A}})-\frac{w_{I_{\text{A}}}}{2}\times\left\langle \left(\left\langle b(s)\right\rangle _{(s|I_{\text{B}})}^{(w_{*})}\right)^{2}\right.\nonumber \\
 &  & -2\left\langle b(s)\right\rangle _{(s|I_{\text{B}})}^{(w_{*})}\left\langle b(s)\right\rangle _{(s|I_{\text{A}})}^{(w_{*})}\label{eq:u_av_av}\\
 &  & \left.+\left\langle b^{2}(s)\right\rangle _{(s|I_{\text{A}})}^{(w_{*})}\right\rangle _{(b|I_{\text{A}})}\nonumber \\
 & =: & V(I_{\text{A}})-\frac{w_{I_{\text{A}}}}{2}\left(\text{I}-2\text{II}+\text{III}\right).
\end{eqnarray}
The three terms $\text{I}$-$\text{III }$occurring therein are
\begin{eqnarray}
\text{\!\!\!\!\!\!\!\!I} & := & \left\langle \left(\left\langle b(s)\right\rangle _{(s|I_{\text{B}})}^{(w_{*})}\right)^{2}\right\rangle _{(b|I_{\text{A}})}\nonumber \\
 & = & \sum_{s,s'\in\mathcal{S}}\mathcal{A}(s|I_{\text{B}})\,\mathcal{A}(s'|I_{\text{B}})\,\left\langle b(s)\,b(s')\right\rangle _{(b|I_{\text{A}})}\!\!\!\!\!\!\!\!\!\!\!\!\!\!\!\!\nonumber \\
 & = & \sum_{s,s'\in\mathcal{S}}\mathcal{A}(s|I_{\text{B}})\,\mathcal{A}(s'|I_{\text{B}})\,\left(\overline{b}^{2}+\gamma^{2}\delta_{s\,s'}\right)\!\!\!\!\!\!\!\!\!\!\!\!\!\!\!\!\nonumber \\
 & = & \overline{b}^{2}+\gamma^{2}\sum_{s\in\mathcal{S}}\,\left[\mathcal{A}(s|I_{\text{B}})\right]^{2}\label{eq:I}\\
\!\!\!\!\!\!\!\!\text{II} & := & \left\langle \left\langle b(s)\right\rangle _{(s|I_{\text{B}})}^{(w_{*})}\left\langle b(s)\right\rangle _{(s|I_{\text{A}})}^{(w_{*})}\right\rangle _{(b|I_{\text{A}})}\nonumber \\
 & = & \sum_{s,s'\in\mathcal{S}}\mathcal{A}(s|I_{\text{B}})\,\mathcal{A}(s'|I_{\text{A}})\,\left(\overline{b}^{2}+\gamma^{2}\delta_{s\,s'}\right)\nonumber \\
 & = & \overline{b}^{2}+\gamma^{2}\sum_{s\in\mathcal{S}}\,\mathcal{A}(s|I_{\text{A}})\,\mathcal{A}(s|I_{\text{B}})\label{eq:II}\\
\!\!\!\!\!\!\!\!\text{III} & := & \left\langle \left\langle b^{2}(s)\right\rangle _{(s|I_{\text{B}})}^{(w_{*})}\right\rangle _{(b|I_{\text{A}})}\nonumber \\
 & = & \sum_{s\in\mathcal{S}}\,\mathcal{A}(s|I_{\text{B}})\left(\overline{b}^{2}+\gamma^{2}\right)\nonumber \\
 & = & \overline{b}^{2}+\gamma^{2},\label{eq:III}
\end{eqnarray}
where we wrote just $\mathcal{A}$ for $\mathcal{A}^{(w_{*})}$ for
brevity. Inserting $\text{I}$-$\text{III}$ into Eq.\ \ref{eq:utility-w-mariginalized}
gives
\begin{eqnarray}
u(I_{\text{B}}) & = & V(I_{\text{A}})-\frac{w_{I_{\text{A}}}}{2}\gamma^{2}\times\nonumber \\
 &  & \left\{ \sum_{s\in\mathcal{S}}\,\left[\mathcal{A}(s|I_{\text{B}})\right]^{2}-2\sum_{s\in\mathcal{S}}\,\mathcal{A}(s|I_{\text{B}})\,\mathcal{A}(s|I_{\text{A}})+1\right\} \nonumber \\
 & = & -\frac{w_{I_{\text{A}}}}{2}\gamma^{2}\sum_{s\in\mathcal{S}}\,\left[\mathcal{A}(s|I_{\text{B}})-\mathcal{A}(s|I_{\text{A}})\right]^{2}+\text{const}(I_{\text{A}})\nonumber \\
\label{eq:utility-w-marginalized-2}
\end{eqnarray}

This expected utility needs to be maximized w.r.t.\ $I_{\text{B}},$
Bob's knowledge after the update. It is obvious that the maximum is
at $I_{\text{B}}=I_{\text{A}}$ as then $\mathcal{A}(s|I_{\text{B}})=\mathcal{A}(s|I_{\text{A}})$
and the negative term becomes zero.

At this maximum we have $\mathcal{P}(s|I_{\text{B}})=\mathcal{P}(s|I_{\text{A}})$
if $w_{*}(s)>0$ for all $s\in\mathcal{S}$, which means that Alice
strives for communicating properly, if possible. Otherwise she tries
to minimizes the $\mathcal{L}^{2}$-norm between her and Bob's attention
distribution functions.

In case $\mathcal{S}$ is continuous, properness appears as well as
Alice's optimal strategy if $G(s,s'):=\overline{b}(s)\overline{b}(s')+D(s,s')$
is either diagonal, $G(s,s')=\sigma_{b}^{2}(s)\,\delta(s-s')$, or
translation invariant, $G(s,s')=G(s-s')$, as is shown in App.\ \ref{sec:Attention-in-Continuous}.
If these conditions are not met, Alice will optimally transmit a biased
attention to Bob.

\section{Relative Attention Entropy\label{sec:Relative-Attention-Entropy}}

\subsection{Derivation}

We have seen how attention appears naturally in a communication scenario
in which an honest sender tries to be supportive to the receiver,
without knowing details of the receiver's utility function except
for having a guess for the variation of its narrowness in different
situations. The measure used by such a sender to match the receiver's
attention function to the own one is then typically of a square distance
form, like in Eq.\ \ref{eq:utility-w-marginalized-2}, maybe with
some bias term as in Eq.\ \ref{eq:biased-utility}. In any way, attention
seems to be a central element of utility aware communication.

This poses the question whether there is a scenario in which relative
attention entropy appears as the measure the sender should use to
choose among possible messages. The answer to this question is yes.

In case Alice assumes that Bob will judge her prediction on the basis
of how much attention was given to the situation that ultimately happened,
and knows the weights Bob will apply to turn probabilities into attentions,
as well as wants to be proper, if possible, the relative attention
entropy can be derived analogously to the derivation of relative entropy
in Sec.\ \ref{subsec:Proper-Coding}, as we see in the following.

Again, we require the measure to be analytical, proper, and calibrated,
and only modify the requirement of locality to \textbf{attention locality}:
Only Bob's attention $\mathcal{A}(s_{*}|I_{\text{B}})$ for the case
$s=s_{*}$ that happens in the end matters for Alice's loss.

Again, at the time Alice has to make her choice, she does not know
which situation $s=s_{*}$ will happen in the end, and therefore needs
to minimize her expected loss
\begin{eqnarray}
\mathcal{L}^{(w)}(I_{\text{A}},I_{\text{B}}) & := & \left\langle \mathcal{L}^{(w)}(s_{*},I_{\text{A}},I_{\text{B}})\right\rangle _{(s_{*}|I_{\text{A}})}
\end{eqnarray}
for deciding which knowledge state Bob $I_{\text{B}}$ should get
(via her message $m$). This loss will depend on the weight function
$w(s)$ that turns probabilities into attentions according to Eq.\ \ref{eq:attention}.
We assume $w(s)>0$ for all $s\in\mathcal{S}$ in the following, such
that $\mathcal{A}(s|I_{\text{A}})=\mathcal{A}(s|I_{\text{B}})$ for
all $s\in\mathcal{S}$ implies $\mathcal{P}(s|I_{\text{A}})=\mathcal{P}(s|I_{\text{B}})$
for all $s\in\mathcal{S}$ and vice versa.

\textbf{Attention locality} implies that $\mathcal{L}^{(w)}(s,I_{\text{A}},I_{\text{B}})$
must depend on $I_{\text{B}}$ through $q(s):=\mathcal{A}(s|I_{\text{B}})$,
whereas the dependence on $I_{\text{A}}$ could still be through $\mathcal{P}(s|I_{\text{A}})$.
However, as the information content of $\mathcal{P}(s|I_{\text{A}})$
and $\mathcal{A}(s|I_{\text{A}})$ are equivalent, it is convenient
to express the dependence on $I_{\text{A}}$ through $p(s):=\mathcal{A}(s|I_{\text{A}})$,
as then properness is given when $q=p$.

Thus we have 
\begin{equation}
\mathcal{L}^{(w)}(s,I_{\text{A}},I_{\text{B}})=\mathcal{L}^{(w)}(s,p(s),q(s))+\lambda\,\left(1-\int\text{d}s'\,q(s')\right),
\end{equation}
where as before we use the function signatures to discriminate different
$\mathcal{L}^{(w)}$'s and introduce a Lagrange multiplier to ensure
that $q(s)$ is normalized.

\textbf{Properness} then requests that the expected loss should be
minimal for $q=p$, implying for all possible $s=s_{*}$

\begin{eqnarray}
0\!\!\!\!\! & =\!\!\!\!\! & \left.\frac{\partial\mathcal{L}^{(w)}(I_{\text{A}},I_{\text{B}})}{\partial q(s_{*})}\right|_{q=p}\nonumber \\
\!\!\!\!\! & =\!\!\!\!\! & \left.\left\langle \frac{\partial}{\partial q(s_{*})}\left[\mathcal{L}^{(w)}(s,p(s),q(s))\!\text{\textasciiacute}-\!\lambda\!\!\int\text{d}s'\,q(s')\right]\right\rangle _{(s|I_{\text{A}})}\right|_{q=p}\nonumber \\
\!\!\!\!\! & =\!\!\!\!\! & \left.\int\text{d}s\,\delta(s-s_{*})\left[\frac{\partial\mathcal{L}^{(w)}(s,p(s),y)}{\partial y}\right]_{y=q(s)}\!\!\!\!\mathcal{P}(s|I_{\text{A}})\right|_{q=p}\!\!\!\!-\lambda\nonumber \\
\!\!\!\!\! & =\!\!\!\!\! & \left.\left[\frac{\partial\mathcal{L}^{(w)}(s,p(s_{*}),y)}{\partial y}\right]_{y=q(s_{*})}\!\!\!\!\frac{\frac{p(s_{*})}{w(s_{*})}}{\int_{\mathcal{S}}ds'\,\frac{p(s')}{w(s')}}\right|_{q=p}\!\!\!\!-\lambda.\nonumber \\
\!\!\!\!\! & =\!\!\!\!\! & \left.\left[\frac{\partial\mathcal{L}^{(w)}(s,x,y)}{\partial y}\right]_{y=x}\frac{x}{w(s_{*})\,\int_{\mathcal{S}}ds'\,\frac{p(s')}{w(s')}}-\lambda\right|_{x=p(s_{*})}\!\!\!\!\!\!\!\!\!\!\!\!\!\!\!\!\!\!.\label{eq:properness-1}
\end{eqnarray}
From this follows
\begin{equation}
\left[\frac{\partial\mathcal{L}^{(w)}(s,x,y)}{\partial y}\right]_{y=x}=\frac{\lambda\,w(s_{*})\,\int_{\mathcal{S}}ds'\,\frac{p(s')}{w(s')}}{x},\label{eq:condition-2}
\end{equation}
which is solved analytically by
\begin{equation}
\mathcal{L}^{(w)}(s,x,y)=\lambda\,w(s)\,\int_{\mathcal{S}}ds'\,\frac{p(s')}{w(s')}\,\ln\,y+c^{(w)}(s,x),\label{eq:consequence-1}
\end{equation}
as can be verified by insertion. We note that 
\begin{eqnarray}
\int_{\mathcal{S}}ds\,\frac{p(s)}{w(s)} & = & \int_{\mathcal{S}}ds\,\frac{w(s)\,\mathcal{P}(s|I_{\text{A}})}{w(s)\int_{\mathcal{S}}ds'w(s')\,\mathcal{P}(s'|I_{\text{A}})}\nonumber \\
 & = & \int_{\mathcal{S}}ds\,\frac{\mathcal{P}(s|I_{\text{A}})}{\int_{\mathcal{S}}ds'w(s')\,\mathcal{P}(s'|I_{\text{A}})}\nonumber \\
 & = & \frac{1}{\int_{\mathcal{S}}ds'w(s')\,\mathcal{P}(s'|I_{\text{A}})}
\end{eqnarray}
and choose $\lambda=1.$

\textbf{Calibration} requests then that $0=\mathcal{L}^{(w)}(s,x,x)$
and therefore 
\begin{equation}
c(s,x)=-\frac{\lambda\,w(s)}{\int_{\mathcal{S}}ds'w(s')\,\mathcal{P}(s'|I_{\text{A}})}\,\ln\,x
\end{equation}
.

Thus, Alice's loss function to choose the message
\begin{eqnarray}
\mathcal{L}^{(w)}(I_{\text{A}},I_{\text{B}}) & = & \int_{\mathcal{S}}ds\,\mathcal{P}(s|I_{\text{A}})\times\nonumber \\
 &  & \left[\frac{w(s)\,\left[\ln\,\mathcal{A}(s|I_{\text{A}})-\ln\,\mathcal{A}(s|I_{\text{B}})\right]}{\int_{\mathcal{S}}ds'w(s')\,\mathcal{P}(s'|I_{\text{A}})}\right]\nonumber \\
 & = & \int_{\mathcal{S}}ds\,\mathcal{A}(s|I_{\text{A}})\,\ln\frac{\mathcal{A}(s|I_{\text{A}})}{\mathcal{A}(s|I_{\text{B}})}\nonumber \\
 & = & \mathcal{D}_{s}^{(w)}(I_{\text{A}},I_{\text{B}})
\end{eqnarray}
turns out to be the relative attention entropy. This closes its derivation.

\subsection{Comparison to Other Scoring Rules\label{subsec:Comparison-to-Other}}

A brief comparison of relative attention entropy to other attention
based score functions is in order.

First we note that in case the weights are constant, $w(s)=\text{const}$,
relative attention entropy reduces to relative entropy.

For the comparison to the communication scenario of Sec.\ \ref{sec:Attention},
in which Alice wants to support Bob as much as possible, but does
only know the curvature of Bob's utility, we investigated the limit
of small relative difference between the attention function, $\Delta(s):=(\mathcal{A}(s|I_{\text{B}})-\mathcal{A}(s|I_{\text{A}}))/\mathcal{A}(s|I_{\text{A}})\ll1$
for all $s\in\mathcal{S}$. In this case, relative attention entropy
is well approximated by
\begin{eqnarray}
\mathcal{D}_{s}^{(w)}(I_{\text{A}},I_{\text{B}}) & \!\!\!\!=\!\!\!\! & -\int_{\mathcal{S}}ds\,\mathcal{A}(s|I_{\text{A}})\,\ln\left[\frac{\mathcal{A}(s|I_{\text{B}})}{\mathcal{A}(s|I_{\text{A}})}\right]\nonumber \\
 & \!\!\!\!=\!\!\!\! & -\int_{\mathcal{S}}ds\,\mathcal{A}(s|I_{\text{A}})\,\ln\left[1+\left(\frac{\mathcal{A}(s|I_{\text{B}})}{\mathcal{A}(s|I_{\text{A}})}-1\right)\right]\nonumber \\
 & \!\!\!\!=\!\!\!\! & -\int_{\mathcal{S}}ds\,\mathcal{A}(s|I_{\text{A}})\,\left[\Delta(s)-\frac{1}{2}\Delta^{2}(s)+\ldots\right]\nonumber \\
 & \!\!\!\!\approx\!\!\!\! & \int_{\mathcal{S}}ds\,\left[\mathcal{A}(s|I_{\text{A}})-\mathcal{A}(s|I_{\text{B}})+\right.\nonumber \\
 &  & \left.\frac{1}{2}\frac{(\mathcal{A}(s|I_{\text{B}})-\mathcal{A}(s|I_{\text{A}}))^{2}}{\mathcal{A}(s|I_{\text{A}})}\right]\nonumber \\
 & \!\!\!\!=\!\!\!\! & \frac{1}{2}\int_{\mathcal{S}}ds\,\frac{(\mathcal{A}(s|I_{\text{B}})-\mathcal{A}(s|I_{\text{A}}))^{2}}{\mathcal{A}(s|I_{\text{A}})},
\end{eqnarray}
which is the well-known information metric. Comparing this to the
negative loss function of Eq.\ \ref{eq:utility-w-marginalized-2},
as generalized to the continuum by Eq.\ \ref{eq:biased-utility}
gives under the assumptions of homogeneity and independence ($G(s,s')\propto\delta(s-s')$,
see App.\ \ref{sec:Attention-in-Continuous} for details)

\begin{equation}
-u(I_{\text{B}})\propto\frac{1}{2}\int_{\mathcal{S}}ds\,(\mathcal{A}(s|I_{\text{B}})-\mathcal{A}(s|I_{\text{A}}))^{2}+\text{const}(I_{\text{A}}).
\end{equation}
There is at least one similarity between these scores, in that deviations
between the attention functions should be avoided as the loss increases
with their square. However, these scores also differ in a significant
point, as for the attention entropy the deviation in attention functions
is reversely weighted with Alice attention. This means that relative
attention entropy allows for larger deviations in regions of higher
attention compared to the utility based score, and smaller deviations
in regions of low attention.

Finally, we note that weighted relative entropy as well as relative
attention entropy are equivalent to scoring rules \cite{landes2013objective,landes2015probabilism}.
Scoring rules evaluate how well a belief $I_{\text{B}}$ matches a
probability $\mathcal{P}(s|I_{\text{A}})$ and are -- in our notation
-- of the functional form
\begin{equation}
S(I_{\text{A}},I_{\text{B}})=\int_{S}ds\,\mathcal{P}(s|I_{\text{A}})\,\mathcal{L}(s,I_{\text{B}}),
\end{equation}
with $\mathcal{L}$ being some loss function that expresses how bad
it is to only believe with the strength $\mathcal{P}(s|I_{\text{B}})$
into an event $s\in\mathcal{S}$ that might happen, with the correct
probability $\mathcal{P}(s|I_{\text{A}})$. Scoring rules are used
to choose the ``best fitting'' belief among a set of beliefs, by
picking the one that has the lowest score. They are called \emph{proper},
if the best fit for $I_{\text{B}}$ is $I_{\text{A}}$ whenever the
latter is part of the set of beliefs to choose $I_{\text{B}}$ from.
Any additive, only $I_{\text{B}}$-independent affine transformations
does not change the minimum of the score. Therefore those lead to
identical results for $I_{\text{B}}$. Thus, we need only to show
for our claim of equivalence that $\widetilde{\mathcal{D}}_{s}^{(w)}(I_{\text{A}},I_{\text{B}})$
and $\mathcal{D}_{s}^{(w)}(I_{\text{A}},I_{\text{B}})$ can be brought
into the extended form

\begin{equation}
S'(I_{\text{A}},I_{\text{B}})=a(I_{\text{A}})\int_{S}ds\,\mathcal{P}(s|I_{\text{A}})\,\mathcal{L}(s,I_{\text{B}})+b(I_{\text{A}}).
\end{equation}

This works for weighted relative entropy by choosing
\begin{eqnarray}
\mathcal{L}(s,I_{\text{B}}) & = & -w(s)\,\ln\mathcal{P}(s|I_{\text{B}})\text{,}\\
a(I_{\text{A}}) & = & 1\text{, and}\\
b(I_{\text{A}}) & = & \int_{\mathcal{S}}ds\,\mathcal{P}(s|I_{\text{A}})\,w(s)\,\ln\mathcal{P}(s|I_{\text{A}}),
\end{eqnarray}
as well as for relative attention entropy with
\begin{eqnarray}
\mathcal{L}(s,I_{\text{B}}) & = & -w(s)\,\ln\mathcal{A}^{(w)}(s|I_{\text{B}})\text{,}\\
a(I_{\text{A}}) & = & \left[\int_{\mathcal{S}}\!ds\,w(s)\,\mathcal{P}(s|I_{\text{A}})\right]^{-1}\text{, and}\\
c(I_{\text{A}}) & = & \int_{\mathcal{S}}ds\,\mathcal{A}^{(w)}(s|I_{\text{A}})\,\ln\mathcal{A}^{(w)}(s|I_{\text{A}}).
\end{eqnarray}

Thus, the well developed formalism of scoring rules \cite{landes2015probabilism}
can be used to investigate these entropies. It might be interesting
to note in that context that weighted entropy is equivalent\footnote{But not identical, due to an irrelevant additional term.}
to a local scoring rule, since its $\mathcal{L}(s,I_{B})$ depends
only on $\mathcal{P}(s|I_{\text{B}})$ for the $s$ in the argument
of $\mathcal{L}$. However, attention entropy is (equivalent to)equivalent
to a non-local score, as the normalization of the attention function
in its $\mathcal{L}$ combines values of $\mathcal{P}(s|I_{\text{B}})$
for different $s$.

\section{Conclusion\label{sec:Conclusion}}

\subsection{Properness, Attention, and Entropy}

Entropy is a central element of communication theory. Relative entropy
allows a sender to decide which message to send in order to inform
about an unknown situation in case only the communicated probability
of the situation that finally happens matters. Naively introducing
an importance weighting for the different situations into relative
entropy renders weighted relative entropy to be improper, meaning
that it does not favor to transmit the sender's precise knowledge
state in case this is possible.

In order to find guidance how a weighting could be introduced into
entropic communication properly, we investigated the scenario in which
a sender, Alice, informs a receiver, Bob, about a situation that will
matter for a decision on an action Bob will perform. The goal of this
exercise is to find a scenario that encourages Alice to be on the
one hand proper, and on the other hand to include weights into her
considerations. Alice can decide which aspects of her knowledge she
communicates and which she omits. In case the utility functions of
Alice and Bob differ, Alice might be tempted to lie to Bob. This would
certainly be improper. We argued that lying should be strongly discouraged
if Alice and Bob interact repeatedly, as otherwise Bob might discover
that Alice lies and stops cooperating with her, or even punishes her
by taking actions that impact her utility negatively. Only the existence
of this option for Bob could give Alice a sufficient incentive towards
honesty.

But even if Alice is bound to be honest, she still can choose what
of her knowledge is revealed to Bob, and what she prefers to keep
to herself by communicating diplomatically. In order to be able to
influence Bob's action to her advance, Alice has to give him some
for him useful information, but only in a way that this information
also serves her interests. This way, both expect to benefit from the
communication, which is honest, but not proper.

Again, in a repeated interaction scenario, Bob has a chance to discover
Alice not being fully supportive to his needs by judging how helpful
Alice's communications were and whether there are systematic omissions
of relevant information. For example, in the scenario discussed in
Sect.\ \ref{subsec:Example-of-Misaligned}, in which Alice's interests
are always rotated by $90{^\circ}$ to the left of Bob's, he might
realize that her advice makes him choose actions that are typically
rotated $45{^\circ}$ to the left of what would have been optimal
for him. Under the plausible assumption that her knowledge is generated
independently from his utility, a few of such incidents should make
him suspicious about Alice really providing him with \textcolor{blue}{all
of her for him relevant information}. Thus, Alice also risks to get
a bad reputation by not being fully supportive to Bob.

Assuming then that Alice aligns her interests with Bob's, we still
do not find that Alice is forced to be proper, as she only needs to
inform him about the aspects of her knowledge that are relevant for
his action.

In order to recover properness in this communication scenario, we
needed to assume that Alice is fully supportive to Bob, but does not
know his utility function in detail. Now she has to inform him properly,
to prepare him for whatever his utility is. Furthermore, if she knows
how sharply his utility function is peaked in the different situations,
she should fold this sharpness as a weight into her measure to choose
how to communicate. More precisely, Alice should turn her knowledge
state into an attention function, basically a weighted probability
distribution that is again normalized to one. And then she should
communicate such that Bob's similarly constructed attention function
becomes as close as possible to hers. In the discussed scenario, the
square difference of the attention functions should be minimized.
This quadratic loss function for attentions has a well known equivalent
for probabilities, the Brier-score \parencite{brier1950verification}.
For this an axiomatic characterization exists \parencite{selten1998axiomatic},
which requests properness as one of the axioms (there called ``incentive
compatibilty''). Here, we found a communication scenario in which
properness emerges from the request that the communication should
be useful for the receiver, without having that use specified.

This last scenario therefore provides a communication measure that
is proper and weighted. It is, however, a quadratic loss and therefore
of a different form than an entropy based on a logarithm. Nevertheless,
it shows the path on how to construct such a weighted entropy that
leads to properness.

In order to have a proper and weighted entropic measure, we have to
request that Alice's communication is judged by Bob on the basis of
which attention value she gave to the situation that finally happened.
This and the request of properness then determines relative attention
entropy as the unique measure for Alice to choose her message.

It should be noted that attention is here formed by giving weights
to different possible situations. In machine learning, the term attention
is prominent in form of weights on different parts of a data vector
or latent space \parencite{chen2015abc,vaswani2017attention,lindsay2020attention}.
These two different concepts of attention are not completely unrelated,
as giving weight to specific parts of the data implies to weight up
possibilities to which these parts of the data point.

Our purely information theoretical motivated considerations should
have technical as well as socio-psychological implications, as we
discuss in the following.

\subsection{Technical Perspective}

The concept of attention and its relative entropy should have a number
of technical applications.

In designing communication systems, the relevance might differ between
situations, about which the communication should inform. Attention
and its relative entropy guide how to incorporate this into the system
design. More specifically, in the problem of Bayesian data compression
one tries to find compressed data that imply an approximate posterior
that is as similar as possible to the original one, which is measured
by their relative entropy \parencite{harth2021toward}. However, there
can be cases in which the relative attention entropy is a better choice
as it permits for importance weighting of the potential situations.

Bayesian updating from a prior $\mathcal{P}(s|I)$ to a posterior
\begin{equation}
\mathcal{P}(s|d,I)=\frac{\mathcal{P}(d|s,I)\,\mathcal{P}(s|I)}{\mathcal{P}(d|I)}=\frac{\mathcal{P}(d|s,I)\,\mathcal{P}(s|I)}{\int ds\,\mathcal{P}(d|s,I)\,\mathcal{P}(s|I)}
\end{equation}
is of the form of forming an attention function out of the prior distribution,
with the weights being given by the likelihood $\mathcal{P}(d|s,I)$.
Communicating a prior in the light of the data one might already have
gotten is then also best done using the corresponding relative attention
entropy.

Furthermore, we like to stress that attention functions as defined
here are formally equivalent to probabilities and can therefore --
formally -- be inserted into any formula that takes those as arguments.
In particular, all scoring rules for probabilities \parencite{gneiting2007strictly}
can be extended to attentions, and therefore attention provides a
mean to introduce the concept of relevance into those.

Finally, we like to point out that ensuring that more relevant dimensions
of a signal or situation $s\in\mathbb{R}^{n}$ are more reliably communicated
can be achieved by constructions like 
\begin{equation}
\widetilde{\mathcal{D}}_{s}^{(w,a)}(I_{\text{A}},I_{\text{B}}):=\mathcal{D}_{s}^{(w)}(I_{\text{A}},I_{\text{B}})+\sum_{i=1}^{n}c_{i}\mathcal{D}_{s_{i}}(I_{\text{A}},I_{\text{B}}),
\end{equation}
in which the additional relative entropies for individual signal directions
are weighted according to $c=(c_{i})_{i=1}^{n}\in\left(\mathbb{R}_{0}^{+}\right)^{n}$.
The term $\mathcal{D}_{s}^{(w)}(I_{\text{A}},I_{\text{B}})$ ensures
propriety of the resulting scoring rule for any $c\in\left(\mathbb{R}_{0}^{+}\right)^{n}$.

\subsection{Socio-psychological Perspective}

Attention, intention, and properness are concepts that play a significant
role in cognition, psychology, and sociology \parencite{neill1987selective,lavie2004load,posner2004attention,chun2011taxonomy,mancas2016human,lindsay2020attention}.
This work made it clear that utility aware communication naturally
involves the concept of attention functions, which guide the choice
of topics to the more important aspects of the speaker's knowledge
that are to be communicated. As there could be certain situations
-- for example -- in which the different options for actions a message
receiver has do not matter much and therefore detailed knowledge of
these situation is not of great value to him. The sender of messages
should not spend much of her valuable communication bandwidth on informing
about these situation of low empowerment to the receiver.

In our derivation of properness and attention we investigated scenarios
in which the interests of speaker and receiver deviated. This is a
very common situation in sociology. We saw that misaligned interests
can leave an imprint in the topic choice of otherwise honest communication
partners. Based on our calculations in Sec.\ \ref{subsec:Misaligned-Interests},
we expect that usefulness of received information decreases the more
the interest of the sender differed from the one of the receiver.
If the interests are exactly oppositely directed, the sender would
prefer to send no information at all. Otherwise, the optimally transmitted
information will result in a compromise between the sender's and the
receiver's interests.

The fact that misalignment of interests in general reduces the information
content of messages in a society of mostly honest actors provides
the possibility to detect and measure the level of such misalignment.
Furthermore, our analysis shows that the specific topic choices made
by communication partners should allow to draw conclusions on their
intentions, and on their believes about the intentions of the receiver
of their messages.

\subsection*{Acknowledgements}

We thank Viktoria Kainz and two anonymous reviewers for constructive
feedback on the manuscript. Philipp Frank acknowledges funding through
the German Federal Ministry of Education and Research for the project
\emph{ErUM-IFT: Informationsfeldtheorie für Experimente an Großforschungsanlagen}
(Förderkennzeichen: 05D23EO1).

\printbibliography

\appendix

\section{Attention Example Calculations \label{sec:Attention-Example-Calculations}}

Here, we give details of the calculation for Sect.\ \ref{subsec:Attention-Example}.
Before we calculate the attention functions we note that weighting
a Gaussian with an exponential weight function $w(s)=\exp(\lambda\,s)$
shifts and re-scales it:
\begin{eqnarray}
 &  & e^{\lambda\,s}\mathcal{G}\left(s-m,\sigma^{2}\right)\\
 & = & \frac{1}{\sqrt{2\pi\sigma^{2}}}\exp\left(\lambda\,s-\frac{(s-m)^{2}}{2\sigma^{2}}\right)\nonumber \\
 & = & \frac{1}{\sqrt{2\pi\sigma^{2}}}\exp\left(-\frac{s^{2}-2s(m+\lambda\,\sigma^{2})+m^{2}}{2\sigma^{2}}\right)\nonumber \\
 & = & \mathcal{G}\left(s-(m+\lambda\,\sigma^{2}),\sigma^{2}\right)\,\exp\left(m\lambda+\frac{1}{2}\lambda^{2}\sigma^{2}\right)\nonumber 
\end{eqnarray}
With this, we see that Alice's attention becomes
\begin{eqnarray*}
\mathcal{A}^{(w)}(s|I_{\text{A}}) & = & \frac{1}{2\mathcal{Z_{\text{A}}}(\lambda)}\sum_{x\in\{-1,1\}}e^{\lambda s}\mathcal{G}\left(s-x,\sigma_{\text{A}}^{2}\right)\\
 & = & \frac{1}{2\mathcal{Z_{\text{A}}}(\lambda)}\sum_{x\in\{-1,1\}}\mathcal{G}\left(s-(x+\lambda\,\sigma_{\text{A}}^{2}),\sigma_{\text{A}}^{2}\right)\\
 &  & \times\exp\left(x\lambda+\frac{1}{2}\lambda^{2}\sigma^{2}\right),
\end{eqnarray*}
with the normalization
\begin{eqnarray}
\mathcal{Z_{\text{A}}}(\lambda) & = & \frac{1}{2}\sum_{x\in\{-1,1\}}\overbrace{\int\text{d}s\,\mathcal{G}\left(s-(x+\lambda\,\sigma_{\text{A}}^{2}),\sigma_{\text{A}}^{2}\right)}^{=1}\nonumber \\
 &  & \times\exp\left(x\lambda+\frac{1}{2}\lambda^{2}\sigma_{\text{A}}^{2}\right)\nonumber \\
 & = & \frac{1}{2}\sum_{x\in\{-1,1\}}\exp\left(x\lambda+\frac{1}{2}\lambda^{2}\sigma_{\text{A}}^{2}\right)\nonumber \\
 & = & \frac{1}{2}\left[e^{\lambda}+e^{-\lambda}\right]e^{\frac{1}{2}\lambda^{2}\sigma_{\text{A}}^{2}}\nonumber \\
 & = & \cosh\lambda\,\exp\left(\frac{1}{2}\lambda^{2}\sigma_{\text{A}}^{2}\right),
\end{eqnarray}
such that 
\begin{eqnarray}
\mathcal{A}^{(w)}(s|I_{\text{A}}) & \!\!\!\!=\!\!\!\! & \frac{1}{2}\!\!\!\!\sum_{x\in\{-1,1\}}\!\!\!\!\frac{e^{x\,\lambda}\mathcal{G}\left(s-x-\lambda\sigma_{\text{A}}^{2},\sigma_{\text{A}}^{2}\right)}{\cosh\lambda},
\end{eqnarray}
as we claimed in Eq.\ \ref{eq:A_A}.

An analogous calculation gives Bob's attention function,
\begin{eqnarray}
\mathcal{A}^{(w)}(s|I_{\text{B}}) & = & \frac{e^{\lambda s}\mathcal{G}\left(s-m,\sigma_{\text{B}}^{2}\right)}{\mathcal{Z}_{\text{B}}(\lambda)},\nonumber \\
 & = & \mathcal{G}\left(s-(m+\lambda\,\sigma_{\text{B}}^{2}),\sigma_{\text{B}}^{2}\right)\times\nonumber \\
 &  & \frac{\exp\left(m\lambda+\frac{1}{2}\lambda^{2}\sigma_{\text{B}}^{2}\right)}{\mathcal{Z}_{\text{B}}(\lambda)}\text{, with}\nonumber \\
\mathcal{Z}_{\text{B}}(\lambda) & = & \int\text{d}s\,\mathcal{G}\left(s-(m+\lambda\,\sigma_{\text{B}}^{2}),\sigma_{\text{B}}^{2}\right)\times\nonumber \\
 &  & \exp\left(m\lambda+\frac{1}{2}\lambda^{2}\sigma_{\text{B}}^{2}\right)\nonumber \\
 & = & \exp\left(m\lambda+\frac{1}{2}\lambda^{2}\sigma_{\text{B}}^{2}\right)\text{ such that}\nonumber \\
\mathcal{A}^{(w)}(s|I_{\text{B}}) & = & \mathcal{G}\left(s-m',\sigma_{\text{B}}^{2}\right),\text{ with}\\
m' & := & m+\lambda\sigma_{\text{B}}^{2},
\end{eqnarray}
as claimed by Eq.\ \ref{eq:A_B}. The relative entropy of these --
up to terms that do not depend on $I_{\text{B}}=(m,\sigma_{\text{B}}^{2})$
and are dropped (as indicated by ``$\widehat{=}$'' in the following
whenever happening) -- is
\begin{eqnarray}
\mathcal{D}_{s}^{(w)}(I_{\text{A}},I_{\text{B}}) & = & \int\text{d}s\,\mathcal{A}^{(w)}(s|I_{\text{A}})\ln\frac{\mathcal{A}^{(w)}(s|I_{\text{A}})}{\mathcal{A}^{(w)}(s|I_{\text{B}})}\nonumber \\
 & \widehat{=} & -\int\text{d}s\,\mathcal{A}^{(w)}(s|I_{\text{A}})\ln\mathcal{A}^{(w)}(s|I_{\text{B}})\nonumber \\
 & \widehat{=} & \int\text{d}s\,\mathcal{A}^{(w)}(s|I_{\text{A}})\frac{1}{2}\left[\frac{\left(s-m'\right)^{2}}{\sigma_{\text{B}}^{2}}+\ln\sigma_{\text{B}}^{2}\right]\nonumber \\
 & = & \frac{1}{2\sigma_{\text{B}}^{2}}\langle\left(s-m'\right)^{2}\rangle_{(s|I_{\text{A}})}^{(w)}+\ln\sigma_{\text{B}},
\end{eqnarray}
where we introduced the attention averaging $\langle f(s)\rangle_{(s|I_{\text{A}})}^{(w)}:=\int\text{d}s\,\mathcal{A}^{(w)}(s|I_{\text{A}})\,f(s)$.
From this it becomes apparent that $\mathcal{A}^{(w)}(s|I_{\text{B}})$
inherits the first and second moment from $\mathcal{A}^{(w)}(s|I_{\text{A}})$
during the minimization of the relative attention entropy w.r.t.\ $I_{\text{B}}$:
\begin{eqnarray}
0 & = & \frac{\partial\mathcal{D}_{s}^{(w)}(I_{\text{A}},I_{\text{B}})}{\partial m'}\nonumber \\
 & = & \frac{1}{\sigma_{\text{B}}^{2}}\langle s-m'\rangle_{(s|I_{\text{A}})}^{(w)}\nonumber \\
\Rightarrow & m'= & \langle s\rangle_{(s|I_{\text{A}})}^{(w)}\\
0 & = & \frac{\partial\mathcal{D}_{s}^{(w)}(I_{\text{A}},I_{\text{B}})}{\partial\sigma_{\text{B}}}\nonumber \\
 & = & -\frac{1}{\sigma_{\text{B}}^{3}}\langle\left(s-m'\right)^{2}\rangle_{(s|I_{\text{A}})}^{(w)}+\frac{1}{\sigma_{\text{B}}}\nonumber \\
\Rightarrow\sigma_{\text{B}}^{2} & = & \langle\left(s-m'\right)^{2}\rangle_{(s|I_{\text{A}})}^{(w)}
\end{eqnarray}
The first moment is 
\begin{eqnarray}
\langle s\rangle_{(s|I_{\text{A}})}^{(w)} & = & \frac{1}{2}\sum_{x\in\{-1,1\}}\frac{e^{x\,\lambda}\left(x+\lambda\sigma_{\text{A}}^{2}\right)}{\cosh\lambda}\nonumber \\
 & = & \frac{e^{\lambda}\left(1+\lambda\sigma_{\text{A}}^{2}\right)-e^{-\lambda}\left(1-\lambda\sigma_{\text{A}}^{2}\right)}{e^{\lambda}+e^{-\lambda}}\nonumber \\
 & = & \tanh\lambda+\lambda\sigma_{\text{A}}^{2}\equiv m'.
\end{eqnarray}
The second moment is 
\begin{eqnarray}
\langle\left(s-m'\right)^{2}\rangle_{(s|I_{\text{A}})}^{(w)} & = & \frac{2}{\cosh\lambda}\sum_{x\in\{-1,1\}}\!\!\!\!e^{x\,\lambda}\times\nonumber \\
 &  & \langle\left(s-m'\right)^{2}\rangle_{\mathcal{G}\left(s-x-\lambda\sigma_{\text{A}}^{2},\sigma_{\text{A}}^{2}\right)}\nonumber \\
 & = & \frac{2}{\cosh\lambda}\sum_{x\in\{-1,1\}}\!\!\!\!e^{x\,\lambda}\times\nonumber \\
 &  & \left[\sigma_{\text{A}}^{2}+\left(m'-x-\lambda\sigma_{\text{A}}^{2}\right)^{2}\right]\nonumber \\
 & = & \frac{2}{\cosh\lambda}\sum_{x\in\{-1,1\}}\!\!\!\!e^{x\,\lambda}\times\nonumber \\
 &  & \left[\sigma_{\text{A}}^{2}+\left(\tanh\lambda-x\right)^{2}\right]\nonumber \\
 & = & \frac{e^{\lambda}\left[\sigma_{\text{A}}^{2}+\left(\tanh\lambda-1\right)^{2}\right]}{e^{\lambda}+e^{-\lambda}}+\nonumber \\
 &  & \frac{e^{-\lambda}\left[\sigma_{\text{A}}^{2}+\left(\tanh\lambda+1\right)^{2}\right]}{e^{\lambda}+e^{-\lambda}}\nonumber \\
 & = & \sigma_{\text{A}}^{2}+\tanh^{2}\lambda+1-2\tanh^{2}\lambda\nonumber \\
 & = & \sigma_{\text{A}}^{2}+1-\tanh^{2}\lambda\equiv\sigma_{\text{B}}^{2},
\end{eqnarray}
as claimed in Eq.\ \ref{eq:sigma_B}. From this it follows that for
the mean of Bob's final knowledge the following equation holds, 
\begin{eqnarray}
m & = & m'-\lambda\sigma_{\text{B}}^{2}\nonumber \\
 & = & \tanh\lambda+\lambda\left(\sigma_{\text{A}}^{2}-\sigma_{\text{B}}^{2}\right)\nonumber \\
 & = & \tanh\lambda-\lambda\left(1-\tanh^{2}\lambda\right)\nonumber \\
 & = & \tanh\lambda-\frac{\lambda}{\cosh^{2}\lambda}
\end{eqnarray}
as claimed by Eq.\ \ref{eq:m}.

Finally, the mean $\widetilde{m}$ and uncertainty dispersion $\widetilde{\sigma}_{\text{B}}^{2}$
of Bob's knowledge state in case Alice uses the weighted relative
entropy of Eq.\ \ref{eq:weighted-relative-entropy} for designing
her message to Bob need to be worked out. This entropy -- up to irrelevant
constant terms -- is 
\begin{eqnarray}
\mathcal{\widetilde{D}}_{s}^{(w)}(I_{\text{A}},\widetilde{I_{\text{B}}}) & \widehat{=} & -\int\text{d}s\,w(s)\,\mathcal{P}(s|I_{\text{A}})\,\ln\mathcal{P}(s|\widetilde{I_{\text{B}}})\nonumber \\
 & = & -\frac{1}{2}\int\text{d}s\,e^{\lambda s}\sum_{x\in\{-1,1\}}\mathcal{G}(s-x,\sigma_{\text{A}}^{2})\times\nonumber \\
 &  & \ln\mathcal{G}(s-\widetilde{m},\widetilde{\sigma}_{\text{B}})\nonumber \\
 & \widehat{=} & \frac{1}{2}\int\text{d}s\sum_{x\in\{-1,1\}}\mathcal{G}\left(s-(x+\lambda\,\sigma_{\text{A}}^{2}),\sigma_{\text{A}}^{2}\right)\times\nonumber \\
 &  & \exp\left(x\lambda+\frac{1}{2}\lambda^{2}\sigma_{\text{A}}^{2}\right)\left[\frac{(s-\widetilde{m})^{2}}{2\widetilde{\sigma}_{\text{B}}^{2}}+\ln\widetilde{\sigma}_{\text{B}}\right]\nonumber \\
 & = & \frac{1}{2}e^{\frac{1}{2}\lambda^{2}\sigma_{\text{A}}^{2}}\sum_{x\in\{-1,1\}}e^{x\lambda}\times\nonumber \\
 &  & \left[\frac{(x+\lambda\,\sigma_{\text{A}}^{2}-\widetilde{m})^{2}+\sigma_{\text{A}}^{2}}{2\widetilde{\sigma}_{\text{B}}^{2}}+\ln\widetilde{\sigma}_{\text{B}}\right].
\end{eqnarray}
Minimizing this w.r.t.\ $\widetilde{m}$ yields
\begin{eqnarray}
0 & = & \frac{\partial\mathcal{\widetilde{D}}_{s}^{(w)}(I_{\text{A}},\widetilde{I_{\text{B}}})}{\partial\widetilde{m}}\nonumber \\
 & = & \frac{1}{2}e^{\frac{1}{2}\lambda^{2}\sigma_{\text{A}}^{2}}\sum_{x\in\{-1,1\}}e^{x\lambda}\frac{x+\lambda\,\sigma_{\text{A}}^{2}-\widetilde{m}}{\widetilde{\sigma}_{\text{B}}^{2}}\nonumber \\
 & \propto & e^{\lambda}\left(1+\lambda\,\sigma_{\text{A}}^{2}-\widetilde{m}\right)+e^{-\lambda}\left(-1+\lambda\,\sigma_{\text{A}}^{2}-\widetilde{m}\right)\nonumber \\
 & \propto & \left(\lambda\,\sigma_{\text{A}}^{2}-\widetilde{m}\right)\cosh\lambda+\sinh\lambda\nonumber \\
 & \propto & \lambda\,\sigma_{\text{A}}^{2}-\widetilde{m}+\tanh\lambda\\
\Rightarrow\widetilde{m} & = & \lambda\,\sigma_{\text{A}}^{2}+\tanh\lambda
\end{eqnarray}
as Eq.\ \ref{eq:m_re} claims. Inserting this into the weighted relative
entropy and minimizing w.r.t.\ $\widetilde{\sigma}_{\text{B}}$ yields

\begin{eqnarray}
0 & = & \frac{\partial\mathcal{\widetilde{D}}_{s}^{(w)}(I_{\text{A}},\widetilde{I_{\text{B}}})}{\partial\widetilde{\sigma}_{\text{B}}}\nonumber \\
 & \propto & \sum_{x\in\{-1,1\}}e^{x\lambda}\times\nonumber \\
 &  & \left[-\frac{(x-\tanh\lambda)^{2}+\sigma_{\text{A}}^{2}}{\widetilde{\sigma}_{\text{B}}^{3}}+\frac{1}{\widetilde{\sigma}_{\text{B}}}\right]\nonumber \\
 & \propto & e^{\lambda}\left[(1-\tanh\lambda)^{2}+\sigma_{\text{A}}^{2}-\widetilde{\sigma}_{\text{B}}^{2}\right]+\nonumber \\
 &  & e^{-\lambda}\left[(1+\tanh\lambda)^{2}+\sigma_{\text{A}}^{2}-\widetilde{\sigma}_{\text{B}}^{2}\right]\nonumber \\
 & \propto & \left(1+\tanh^{2}\lambda+\sigma_{\text{A}}^{2}-\widetilde{\sigma}_{\text{B}}^{2}\right)\cosh\lambda+\nonumber \\
 &  & -\tanh\lambda\sinh\lambda\nonumber \\
 & \propto & 1+\tanh^{2}\lambda+\sigma_{\text{A}}^{2}-\widetilde{\sigma}_{\text{B}}^{2}-\tanh^{2}\lambda\nonumber \\
 & = & 1+\sigma_{\text{A}}^{2}-\widetilde{\sigma}_{\text{B}}^{2}\nonumber \\
\Rightarrow\widetilde{\sigma}_{\text{B}}^{2} & = & 1+\sigma_{\text{A}}^{2},
\end{eqnarray}
which was claimed by Eq.\ \ref{eq:sigma_re}. This completes the
calculations for Sect.\ \ref{subsec:Attention-Example}.

\section{Real World Communication\label{sec:Real-World-Communication}}

We want to illustrate how Alice's moment constraining messages of
the form of Eq.\ \ref{eq:statement} can embrace ordinary, real world
communications with an example. A general proof that the communication
of moments are sufficiently rich to express any message is beyond
the scope of this work.

To have an illustrative example, we look at the statement $m=\text{\textquotedblleft\text{T}omorrow's weather should be alright\textquotedblright}$.
The relevant, but unknown situation $s$ is tomorrow's weather, which
we assume for the sake of the argument to be out of $\mathcal{S}=\{\text{bad, alright, good}\}\equiv\{-1,0,1\}$,
the latter being a numerical embedding of these situations. It is
reasonable then to assume that the statement contains the message
$``\langle s\rangle_{(s|I_{A})}=0"$, i.e. the first components of
$(f,d)$ are $f_{1}(s)=s$ and $d_{1}=0$. Furthermore the word $\text{\text{\textquotedblleft}should\text{\textquotedblright}}\in\{\text{\text{\textquotedblleft}is going to\text{\textquotedblright}},\text{\text{\textquotedblleft}should\text{\textquotedblright}},\text{\text{\textquotedblleft}might\text{\textquotedblright}}\}\equiv\{0,\nicefrac{1}{4},\nicefrac{1}{2}\}$
can be read as a quantifier for the sender's uncertainty on the situation,
which shall here be interpreted as a statement on the variance, $``\langle(s-d_{1})^{2}\rangle_{(s|I_{A})}=\nicefrac{1}{4}"$,
implying $f_{2}(s)=(s-d_{1})^{2}=s^{2}$ and $d_{2}=\nicefrac{1}{4}$.
Thus the message is given by $m=(f(s),d)=((s,s^{2})^{\text{t}},(0,\nicefrac{1}{4})^{\text{t}}).$

As no prior information is specified we assume $q(s)=\mathcal{P}(s|I_{B}')=\nicefrac{1}{3}$.
This leads to $\mathcal{Z}(\mu)=\frac{1}{3}\sum_{s=-1}^{1}\exp(\mu_{1}\,s+\mu_{2}\,s^{2})$
and therefore to
\begin{eqnarray}
d & = & \begin{pmatrix}0\\
\nicefrac{1}{4}
\end{pmatrix}\overset{!}{=}\frac{\partial\ln\mathcal{Z}(\mu)}{\partial\mu}\nonumber \\
 & = & \frac{\nicefrac{1}{3}}{\mathcal{Z}(\mu)}\sum_{s=-1}^{1}\begin{pmatrix}s\\
s^{2}
\end{pmatrix}\exp(\mu_{1}\,s+\mu_{2}\,(s-d_{1})^{2})\nonumber \\
 & = & \frac{1}{e^{-\mu_{1}+\mu_{2}}+1+e^{\mu_{1}+\mu_{2}}}\begin{pmatrix}-e^{-\mu_{1}+\mu_{2}}+e^{\mu_{1}+\mu_{2}}\\
e^{-\mu_{1}+\mu_{2}}+e^{\mu_{1}+\mu_{2}}
\end{pmatrix}\nonumber \\
 & = & \frac{1}{e^{-\mu_{2}}+2\cosh\mu_{1}}\begin{pmatrix}2\sinh\mu_{1}\\
2\cosh\mu_{1}
\end{pmatrix},
\end{eqnarray}
from which it follows that $\mu_{1}=0$, $\cosh\mu_{1}=1$, and therefore
$\mu_{2}=-\ln\left(6\right)$. Given that then $\mathcal{Z}(\mu)=\frac{1}{3}\,(1+2e^{\mu_{2}})=\nicefrac{4}{9},$
we get as our weather prognosis $\mathcal{P}(\text{bad}|m)=\mathcal{P}(\text{good}|m)=\frac{1}{\mathcal{Z}(\mu)}\times q(s)\times e^{-\mu_{2}s^{2}}=\frac{9}{4}\times\frac{1}{3}\times\frac{1}{6}=\frac{1}{8}$
and $\mathcal{P}(\text{alright}|m)=\frac{9}{4}\times\frac{1}{3}\times1=\frac{3}{4}$.

The alternative statement $m'=\text{\textquotedblleft\text{T}omorrow's weather might be alright\textquotedblright}$
implies $d_{2}=\nicefrac{1}{2}$, $\mu_{2}=-\ln\left(2\right)$, $\mathcal{Z}(\mu)=\frac{1}{3}\,(1+2e^{\mu_{2}})=\nicefrac{2}{3}$,
and therefore leads to a bit less confidence on tomorrow's weather
with $\mathcal{P}(\text{alright}|m')=\nicefrac{1}{2}$ and $\mathcal{P}(\text{bad}|m')=\mathcal{P}(\text{good}|m')=\nicefrac{1}{4}$.

Of course the here chosen language embedding -- meaning a representation
of a language in a mathematical structure, as here the representation
of statements on weather in terms of topic function $f$ and message
data $d$ -- is only one possibility out of many. The language embedding
used by the speaker and recipient of a message needs to be identical
for a high fidelity communication. In reality, the embedding will
depend on social conventions that can differ between speaker and recipient.
This might in part explain the difficulty of communication across
cultures, even if a common language is used.

\section{Accurate Communication\label{sec:Accurate-Communication}}

Here, we show that the message format of Eq.\ \ref{eq:statement}
permits Alice in principle to transfer her exact knowledge to Bob
if there are no bandwidth restrictions. In case she knows his knowledge
state $\mathcal{P}(s|I_{0})$, she could simply send the relative
surprise function $f(s)=-\ln(\mathcal{P}(s|I_{\text{A}})/\mathcal{P}(s|I_{0}))$
as well as $d=\langle f(s)\rangle_{(s|I_{\text{A}})}=\mathcal{D}_{s}(I_{\text{A}},I_{0})$,
which turns out to be the amount of information she is transmitting.
Bob updates then to $\mathcal{P}(s|I_{\text{B}})=\mathcal{P}(s|I_{\text{A}})$
as a straightforward calculation shows: 
\begin{eqnarray}
\mathcal{P}(s|I_{B}) & = & \frac{\mathcal{P}(s|I_{0})}{\mathcal{Z}(\mu)}\,\exp\left(-\mu\ln\frac{\mathcal{P}(s|I_{\text{A}})}{\mathcal{P}(s|I_{0})}\right)\nonumber \\
 & = & \frac{\mathcal{P}(s|I_{0})}{\mathcal{Z}(\mu)}\,\left[\frac{\mathcal{P}(s|I_{\text{A}})}{\mathcal{P}(s|I_{0})}\right]^{-\mu}\\
\mathcal{Z}(\mu) & = & \int\text{d}s\,\mathcal{P}(s|I_{0})\,\left[\frac{\mathcal{P}(s|I_{\text{A}})}{\mathcal{P}(s|I_{0})}\right]^{-\mu}\\
d & = & \mathcal{D}_{s}(I_{\text{A}},I_{0})\overset{!}{=}\frac{\partial\ln\mathcal{Z}(\mu)}{\partial\mu}\\
 & = & \frac{-\int\text{d}s\,\mathcal{P}(s|I_{0})\,\left[\frac{\mathcal{P}(s|I_{\text{A}})}{\mathcal{P}(s|I_{0})}\right]^{-\mu}\ln\left(\frac{\mathcal{P}(s|I_{\text{A}})}{\mathcal{P}(s|I_{0})}\right)}{\int\text{d}s\,\mathcal{P}(s|I_{0})\,\left[\frac{\mathcal{P}(s|I_{\text{A}})}{\mathcal{P}(s|I_{0})}\right]^{-\mu}}\nonumber \\
 & \overset{\!\!\!\!\mu=-1\!\!\!\!}{=} & \frac{-\int\text{d}s\,\mathcal{P}(s|I_{\text{A}})\ln\left(\frac{\mathcal{P}(s|I_{\text{A}})}{\mathcal{P}(s|I_{0})}\right)}{\int\text{d}s\,\mathcal{P}(s|I_{\text{A}})}=\mathcal{D}_{s}(I_{\text{A}},I_{0})\nonumber \\
\Rightarrow\mu & = & -1\;\;\;\;\Rightarrow\\
\mathcal{P}(s|I_{\text{B}}) & = & \mathcal{P}(s|I_{\text{A}})
\end{eqnarray}
In case she does not know his initial belief, she could alternatively
send her knowledge by using the vector valued topic $f(s)=(\delta(s-s'))_{s'\in\mathcal{S}}$.
This lets the message data $d=\int\text{d}s\,\mathcal{P}(s|I_{\text{A}})\,(\delta(s-s'))_{s'\in\mathcal{S}}=(\mathcal{P}(s'|I_{\text{A}}))_{s'\in\mathcal{S}}$
be a vector that contains her full probability function to which Bob
would then update his knowledge.

\section{Topic Gradient\label{sec:Topic-Gradient}}

Here we work out the gradient of Alice's utility w.r.t. the topic
of her honest communication given in Sect.\ \ref{subsec:Optimal-Action}
according to Eqs.\ \ref{eq:uA(f)}-\ref{eq:mu}. This gradient
\begin{eqnarray}
\frac{\text{d}u_{\text{A}}(f)}{\text{d}f(s_{*})} & = & \left\langle \frac{\text{d}u_{\text{A}}(s,a_{\text{B}})}{\text{d}f(s_{*})}\right\rangle _{(s|I_{\text{A}})}\nonumber \\
 & = & \left\langle \left(\frac{\partial u_{\text{A}}(s,a_{\text{B}})}{\partial a_{\text{B}}}\right)^{\text{t}}\,\frac{\text{d}a_{\text{B}}}{\text{d}f(s_{*})}\right\rangle _{(s|I_{\text{A}})}\nonumber \\
 & = & \left\langle g_{\text{A}}(s,a_{\text{B}})\right\rangle _{(s|I_{\text{A}})}^{\text{t}}\frac{\text{d}a_{\text{B}}}{\text{d}f(s_{*})}
\end{eqnarray}
consists of a product between Alice's expectation for her utility
gradient $\left\langle g_{\text{A}}(s,a_{\text{B}})\right\rangle _{(s|I_{\text{A}})}$
given Bob's action $a_{\text{B}}$ and how his action changes with
changing topics of her communication. This gradient vanishes when
Bob happens to choose the for Alice optimal action, such that $\left\langle g_{\text{A}}(s,a_{\text{B}})\right\rangle _{(s|I_{\text{A}})}=0$
and therefore $\left\langle g_{\text{A}}(s,a_{\text{B}})\right\rangle _{(s|I_{\text{A}})}=\left\langle g_{\text{B}}(s,a_{\text{B}})\right\rangle _{(s|I_{\text{B}})}$
as the latter is zero thanks to Bob's choice of action, see Eq.\ \ref{eq:uB-optimal},
or when a further change in $f(s_{*})$ does not change Bob's action
$a_{\text{B}}$ any more.

Bob's chosen action is the result of the minimization in Eq.\ \ref{eq:aB}.
Its gradient w.r.t.\ $f$ can be worked out using the implicit function
theorem: 
\begin{eqnarray}
\frac{\text{d}a_{\text{B}}}{\text{d}f(s_{*})} & \!\!\!\!=\!\!\!\! & -\left[\frac{\partial^{2}u_{\text{B}}(a_{\text{B}}|I_{\text{B}})}{\partial a_{\text{B}}\partial a_{\text{B}}^{\text{t}}}\right]^{-1}\,\frac{\partial\,\text{d}u_{\text{B}}(a_{\text{B}}|I_{\text{B}})}{\partial a_{\text{B}}\,\text{d}f(s_{*})}.
\end{eqnarray}

The last term of this is
\begin{eqnarray}
\frac{\partial\,\text{d}u_{\text{B}}(a|I_{\text{B}})}{\partial a\,\text{d}f(s_{*})}\!\!\!\! & = & \!\!\!\!\frac{\text{d}}{\text{d}f(s_{*})}\left\langle \frac{\partial u_{\text{B}}(s,a)}{\partial a}\right\rangle _{(s|I_{\text{B}})}\nonumber \\
 & = & \!\!\!\!\frac{\text{d}}{\text{d}f(s_{*})}\left\langle g_{\text{B}}(s,a)\right\rangle _{(s|I_{\text{B}})}\nonumber \\
 & = & \!\!\!\!\int\text{d}s\,g_{\text{B}}(s,a)\,\frac{\text{d}\mathcal{P}(s|I_{\text{B}})}{\text{d}f(s_{*})}\nonumber \\
 & = & \!\!\!\!\int\text{d}s\,g_{\text{B}}(s,a)\,\mathcal{P}(s|I_{\text{B}})\frac{\text{d}\mathcal{\ln P}(s|I_{\text{B}})}{\text{d}f(s_{*})}\nonumber \\
 & = & \!\!\!\!\left\langle g_{\text{B}}(s,a)\frac{\text{d}\left[\mu^{\text{t}}f(s)-\mathcal{\ln Z}(\mu,f)\right]}{\text{d}f(s_{*})}\right\rangle _{(s|I_{\text{B}})}\nonumber \\
 & = & \!\!\!\!g_{\text{B}}(s_{*},a)\,\mathcal{P}(s_{*}|I_{\text{B}})\,\mu^{\text{t}}\nonumber \\
 &  & \!\!\!\!+\left\langle g_{\text{B}}(s,a)f(s)^{\text{t}}\right\rangle _{(s|I_{\text{B}})}\frac{\text{d}\mu}{\text{d}f(s_{*})}\nonumber \\
 &  & \!\!\!\!-\underbrace{\left\langle g_{\text{B}}(s,a)\right\rangle }_{=0}\frac{\text{d}\ln\mathcal{Z}(\mu,f)}{\text{d}f(s_{*})}\nonumber \\
 & = & \!\!\!\!g_{\text{B}}(s_{*},a)\,\mathcal{P}(s_{*}|I_{\text{B}})\,\mu^{\text{t}}\nonumber \\
 &  & \!\!\!\!-\left\langle g_{\text{B}}(s,a)\delta f(s)^{\text{t}}\right\rangle _{(s|I_{\text{B}})}\left\langle \delta f(s)\,\delta f(s)^{\text{t}}\right\rangle _{(s|I_{\text{B}})}^{-1}\nonumber \\
 &  & \!\!\!\!\times\left[\mathcal{P}(s_{*}|I_{\text{B}})-\mathcal{P}(s_{*}|I_{\text{A}})\right],
\end{eqnarray}
with 
\begin{eqnarray}
\delta f(s) & := & f(s)-\left\langle f(s)\right\rangle _{(s|I_{\text{B}})},
\end{eqnarray}
since according to the inverse function theorem applied to the quantity
determining $\mu$
\begin{equation}
\mu:0=g(\mu,f):=\frac{\partial\mathcal{\ln Z}(\mu,f)}{\partial\mu}-\left\langle f(s)\right\rangle _{(s|I_{\text{A}})}
\end{equation}
we have
\begin{eqnarray}
\frac{\text{d}\mu}{\text{d}f(s_{*})} & = & -\left[\frac{\partial}{\partial\mu}g(\mu,f)^{\text{t}}\right]^{-1}\frac{\partial g(\mu,f)}{\partial f(s_{*})}\\
 & = & -\left[\frac{\partial}{\partial\mu}\left(\frac{\partial\mathcal{\ln Z}(\mu,f)}{\partial\mu}-\left\langle f(s)\right\rangle _{(s|I_{\text{A}})}\right)^{\text{t}}\right]^{-1}\nonumber \\
 &  & \times\left[\frac{\partial}{\partial f(s_{*})}\left(\frac{\partial\mathcal{\ln Z}(\mu,f)}{\partial\mu}-\left\langle f(s)\right\rangle _{(s|I_{\text{A}})}\right)\right]\nonumber \\
 & = & -\left\langle \delta f(s)\,\delta f(s)^{\text{t}}\right\rangle _{(s|I_{\text{B}})}^{-1}\nonumber \\
 &  & \left[\frac{\partial}{\partial f(s_{*})}\left(\left\langle f(s)\right\rangle _{(s|I_{\text{B}})}-\left\langle f(s)\right\rangle _{(s|I_{\text{A}})}\right)\right]\nonumber \\
 & = & -\left\langle \delta f(s)\,\delta f(s)^{\text{t}}\right\rangle _{(s|I_{\text{B}})}^{-1}\,\left[\mathcal{P}(s_{*}|I_{\text{B}})-\mathcal{P}(s_{*}|I_{\text{A}})\right]\nonumber 
\end{eqnarray}
thanks to
\begin{eqnarray}
\frac{\partial^{2}\ln\mathcal{Z}(\mu,f)}{\partial\mu\partial\mu^{\text{t}}}\!\!\!\! & = & \!\!\!\!\frac{\partial}{\partial\mu}\left(\frac{\int\text{d}s\,q(s)\,f(s)\,\exp(\mu^{\text{t}}f(s))}{\mathcal{Z}(\mu,f)}\right)\nonumber \\
 & = & \!\!\!\!\frac{\int\text{d}s\,q(s)\,f(s)\,f(s)^{\text{t}}\exp(\mu^{\text{t}}f(s))}{\mathcal{Z}(\mu,f)}\nonumber \\
 &  & \!\!\!\!-\left[\frac{\int\text{d}s\,q(s)\,f(s)\,\exp(\mu^{\text{t}}f(s))}{\mathcal{Z}(\mu,f)}\right]\nonumber \\
 &  & \!\!\!\!\times\left[\frac{\int\text{d}s\,q(s)\,f(s)\,\exp(\mu^{\text{t}}f(s))}{\mathcal{Z}(\mu,f)}\right]^{\text{t}}\nonumber \\
 & = & \!\!\!\!\left\langle f(s)f(s)^{\text{t}}\right\rangle _{(s|I_{\text{B}})}-\left\langle f(s)\right\rangle _{(s|I_{\text{B}})}\!\left\langle f(s)^{\text{t}}\right\rangle _{(s|I_{\text{B}})}\nonumber \\
 & = & \!\!\!\!\left\langle \delta f(s)\,\delta f(s)^{\text{t}}\right\rangle _{(s|I_{\text{B}})}.
\end{eqnarray}
Collecting terms gives the topic gradient of Alice's utility as
\begin{eqnarray}
\frac{\text{d}u_{\text{A}}(f)}{\text{d}f(s_{*})}\!\!\!\! & = & \!\!\!\!-\left\langle g_{\text{A}}(s,a)\right\rangle _{(s|I_{\text{A}})}^{\text{t}}\left[\frac{\partial^{2}u_{\text{B}}(a|I_{\text{B}})}{\partial a\partial a^{\text{t}}}\right]^{-1}\times\label{eq:topicgradient}\\
 &  & \!\!\!\!\left\{ g_{\text{B}}(s_{*},a)\,\mathcal{P}(s_{*}|I_{\text{B}})\,\mu^{\text{t}}-\left\langle g_{\text{B}}(s,a)\delta f(s)^{\text{t}}\right\rangle _{(s|I_{\text{B}})}\right.\nonumber \\
 &  & \!\!\!\!\left.\times\left\langle \delta f(s)\,\delta f(s)^{\text{t}}\right\rangle _{(s|I_{\text{B}})}^{-1}\left[\mathcal{P}(s_{*}|I_{\text{B}})-\mathcal{P}(s_{*}|I_{\text{A}})\right]\right\} ,\nonumber 
\end{eqnarray}
with $a=a_{\text{B }}$ and $\mu$ according to Eq.\ \ref{eq:aB}
and \ref{eq:mu}, respectively.

\section{Specific Topic Gradient\label{sec:Specific-topic-gradient}}

Here, the topic gradient given by Eq.\ \ref{eq:topic_gradient} at
$f(s)=\tau^{\text{t}}s$ is calculated for the simple example of misaligned
interests of Alice and Bob as discussed in Sec.\ \ref{subsec:Misaligned-Interests}.

In order to have a concise notation, let us first note that
\begin{eqnarray}
a & = & a_{\text{B}}=\tau\,\tau^{\text{t}}\overline{s}_{\text{A}}=\tau\,\cos\alpha,\\
\tau & = & \begin{pmatrix}\cos\alpha\\
\sin\alpha
\end{pmatrix},\\
a_{\text{A}} & = & \begin{pmatrix}\cos\varphi\\
\sin\varphi
\end{pmatrix},\\
\mu & = & \tau^{\text{t}}\overline{s}_{\text{A}}=\cos\alpha,
\end{eqnarray}
and therefore 
\begin{equation}
a=\mu\,\tau.
\end{equation}
With this, the building blocks of the gradient given by Eq.\ \ref{eq:topic_gradient}
are
\begin{eqnarray}
\left\langle g_{\text{A}}(s,a)\right\rangle _{(s|I_{\text{A}})}\!\!\!\! & = & \!\!\!\!\left\langle \frac{\partial u_{\text{A}}(s,a)}{\partial a}\right\rangle _{(s|I_{\text{A}})}\nonumber \\
 & = & \!\!\!\!2\left\langle R\,s-a\right\rangle _{(s|I_{\text{A}})}=2\left(a_{\text{A}}-a\right)\\
\left[\frac{\partial^{2}u_{\text{B}}(a|I_{\text{B}})}{\partial a\partial a^{\text{t}}}\right]^{-1}\!\!\!\! & = & \!\!\!\!\left[2\mathbb{1}\right]^{-1}=\frac{1}{2}\mathbb{1},\\
g_{\text{B}}(s_{*},a)\!\!\!\! & = & \!\!\!\!\frac{\partial u_{\text{A}}(s_{*},a)}{\partial a}=2\left(s_{*}-a\right)\text{,}\\
\mathcal{P}(s_{*}|I_{\text{B}})\!\!\!\! & = & \!\!\!\!\mathcal{G}(s-a,\mathbb{1})\nonumber \\
\left\langle g_{\text{B}}(s,a)\delta f(s)^{\text{t}}\right\rangle _{(s|I_{\text{B}})}\!\!\!\! & = & \!\!\!\!2\left\langle \left(s-\tau\,\tau^{\text{t}}\overline{s}_{\text{A}}\right)\left(\tau^{\text{t}}s-\tau^{\text{t}}\overline{s}_{\text{A}}\right)^{\text{t}}\right\rangle _{(s|I_{\text{B}})}\nonumber \\
 & = & \!\!\!\!2\left\langle \left(s-\tau\,\tau^{\text{t}}\overline{s}_{\text{A}}\right)\left(s-\overline{s}_{\text{A}}\right)^{\text{t}}\right\rangle _{(s|I_{\text{B}})}\tau\nonumber \\
 & = & \!\!\!\!2\left[\left\langle ss^{\text{t}}\right\rangle _{(s|I_{\text{B}})}-\left\langle s\right\rangle _{(s|I_{\text{B}})}\overline{s}_{\text{A}}^{\text{t}}\right.\nonumber \\
 &  & \!\!\!\!\left.-\tau\,\tau^{\text{t}}\overline{s}_{\text{A}}\left\langle s^{\text{t}}\right\rangle _{(s|I_{\text{B}})}+\tau\,\tau^{\text{t}}\overline{s}_{\text{A}}\overline{s}_{\text{A}}^{\text{t}}\right]\tau\nonumber \\
 & = & \!\!\!\!2\left[\mathbb{1}-0-0+\left(\tau^{\text{t}}\overline{s}_{\text{A}}\right)^{2}\right]\,\tau\nonumber \\
 & = & \!\!\!\!2\left(1+\cos^{2}\alpha\right)\,\tau\nonumber \\
 & = & \!\!\!\!2\left(1+\mu^{2}\right)\,\tau,\\
\left\langle \delta f(s)\,\delta f(s)^{\text{t}}\right\rangle _{(s|I_{\text{B}})}\!\!\!\! & = & \!\!\!\!\left\langle \left(\tau^{\text{t}}s-\tau^{\text{t}}\overline{s}_{\text{A}}\right)\left(\tau^{\text{t}}s-\tau^{\text{t}}\overline{s}_{\text{A}}\right)^{\text{t}}\right\rangle _{(s|I_{\text{B}})}\nonumber \\
 & = & \!\!\!\!\tau^{\text{t}}\left\langle \left(s-\overline{s}_{\text{A}}\right)\left(s-\overline{s}_{\text{A}}\right)^{\text{t}}\right\rangle _{(s|I_{\text{B}})}\tau\nonumber \\
 & = & \!\!\!\!\tau^{\text{t}}\left[\left\langle ss^{\text{t}}\right\rangle _{(s|I_{\text{B}})}-\left\langle s\right\rangle _{(s|I_{\text{B}})}\overline{s}_{\text{A}}^{\text{t}}\right.\nonumber \\
 &  & \!\!\!\!\left.-\overline{s}_{\text{A}}\left\langle s^{\text{t}}\right\rangle _{(s|I_{\text{B}})}+\overline{s}_{\text{A}}\overline{s}_{\text{A}}^{\text{t}}\right]\tau\nonumber \\
 & = & \!\!\!\!\tau^{\text{t}}\mathbb{1}\tau-0-0+\tau^{\text{t}}\overline{s}_{\text{A}}\,\overline{s}_{\text{A}}^{\text{t}}\tau\nonumber \\
 & = & \!\!\!\!1+\cos^{2}\alpha\nonumber \\
 & = & \!\!\!\!1+\mu^{2},\\
\mathcal{P}(s_{*}|I_{\text{A}}) & = & \!\!\!\!\mathcal{G}(s-\overline{s}_{\text{A}},S_{\text{A}}).
\end{eqnarray}
Inserting these terms into the topic gradient of Alice's utility yields
\begin{eqnarray}
\frac{\text{d}u_{\text{A}}(f)}{\text{d}f(s_{*})}\!\!\!\! & = & \!\!\!\!-\left\langle g_{\text{A}}(s,a)\right\rangle _{(s|I_{\text{A}})}^{\text{t}}\left[\frac{\partial^{2}u_{\text{B}}(a|I_{\text{B}})}{\partial a\partial a^{\text{t}}}\right]^{-1}\times\nonumber \\
 &  & \!\!\!\!\left\{ g_{\text{B}}(s_{*},a)\,\mathcal{P}(s_{*}|I_{\text{B}})\,\mu^{\text{t}}-\left\langle g_{\text{B}}(s,a)\delta f(s)^{\text{t}}\right\rangle _{(s|I_{\text{B}})}\right.\nonumber \\
 &  & \!\!\!\!\left.\times\left\langle \delta f(s)\,\delta f(s)^{\text{t}}\right\rangle _{(s|I_{\text{B}})}^{-1}\left[\mathcal{P}(s_{*}|I_{\text{B}})-\mathcal{P}(s_{*}|I_{\text{A}})\right]\right\} ,\nonumber \\
 & = & \!\!\!\!-2\left(a_{\text{A}}-a\right)^{\text{t}}\frac{1}{2}\mathbb{1}\nonumber \\
 &  & \!\!\!\!\left\{ 2\left(s_{*}-a\right)\,\mathcal{G}(s-a,\mathbb{1})\,\mu-2\left(1+\mu^{2}\right)\,\tau\right.\nonumber \\
 &  & \!\!\!\!\left.\times\left(1+\mu^{2}\right)\left[\mathcal{G}(s-a,\mathbb{1})-\mathcal{G}(s-a_{\text{A}},S_{\text{A}})\right]\right\} ,\nonumber \\
 & = & \!\!\!\!-2\left(a_{\text{A}}-a\right)^{\text{t}}\left\{ \left[\left(s_{*}-a\right)\mu-\tau\right]\,\mathcal{G}(s-a,\mathbb{1})\right.\nonumber \\
 &  & \!\!\!\!\left.+\tau\,\mathcal{G}(s-a_{\text{A}},S_{\text{A}})\right\} ,
\end{eqnarray}
This obviously does not vanish for all $s_{*}$ unless $a_{\text{A}}=a$
or simultaneously $\tau\perp(a_{\text{A}}-a)$ and $\mu=0$, where
the latter means that Alice's message contained no news for Bob. For
example for $\varphi=\nicefrac{\pi}{2}$, where $a_{\text{A}}=(0,1)^{\text{t}}$,
$\tau=\nicefrac{1}{\sqrt{2}}(1,1)^{\text{t}}$, $a=a_{\text{B}}=(\nicefrac{1}{2},\nicefrac{1}{2})^{\text{t}}$,
and $\mu=\nicefrac{1}{\sqrt{2}}$ the gradient is 
\begin{eqnarray}
\frac{\text{d}u_{\text{A}}(f)}{\text{d}f(s_{*})}\!\!\!\! & = & \!\!\!\!-(1,-1)\left\{ \left[\nicefrac{1}{\sqrt{2}}\left(s_{*}-(\nicefrac{3}{2},\nicefrac{3}{2})^{\text{t}}\right)\right]\,\times\right.\nonumber \\
 &  & \mathcal{G}(s_{*}-(\nicefrac{1}{2},\nicefrac{1}{2})^{\text{t}},\mathbb{1})\nonumber \\
 &  & \!\!\!\!\left.+\nicefrac{1}{\sqrt{2}}(1,1)^{\text{t}}\,\mathcal{G}(s-(0,1)^{\text{t}},S_{\text{A}})\right\} \nonumber \\
 & = & \!\!\!\!\nicefrac{1}{\sqrt{2}}\left[(-1,1)\,s_{*}\right]\,\mathcal{P}(s|I_{\text{B}}).\label{eq:specific_gradient_for_orthogonal_interests}
\end{eqnarray}

\section{Attention in Continuous Situations\label{sec:Attention-in-Continuous}}

Here we repeat the calculation from Sec.\ \ref{sec:Relative-Attention-Entropy}
for the case of a continuous situation space $\mathcal{S}$ and potentially
inhomogeneous and correlated knowledge of Alice on $b(s)$.

For this calculation, it is convenient to switch to an information
field theoretical notation \parencite{2009PhRvD..80j5005E,https://doi.org/10.1002/andp.201800127}
by defining the fields $b:=b(\cdot)$, $\mathcal{A_{\text{A}}}:=\mathcal{A}(\cdot|I_{\text{A}})$,
$\mathcal{A_{\text{B}}:=\mathcal{A}}(\cdot|I_{\text{B}})$, and their
scalar product $b^{\dagger}\mathcal{A_{\text{A}}}:=\int_{\mathcal{S}}ds\,b(s)\mathcal{A}(s|I_{\text{A}})$.
Alice might not know $b$ but only have a vague idea about it, which
is characterized by a mean and uncertainty covariance

\begin{eqnarray}
b_{\text{A}} & := & \left\langle b\right\rangle _{\!(b|I_{\text{A}})},\\
D & := & \left\langle \left(b-\overline{b}\right)\,\left(b-\overline{b}\right)^{\dagger}\right\rangle _{\!(b|I_{\text{A}})}.
\end{eqnarray}

For the latter usage we note that
\begin{equation}
\left\langle b\,b^{\dagger}\right\rangle _{(b|I_{\text{A}})}=b_{\text{A}}\,b_{\text{A}}^{\dagger}+D=:G.
\end{equation}

The expected utility is still given by Eq.\ \ref{eq:u_av_av} with
the three terms occurring therein being now
\begin{eqnarray}
\text{\!\!\!\!\!\!\!\!I} & := & \left\langle \left(\left\langle b(s)\right\rangle _{(s|I_{\text{B}})}^{(w_{*})}\right)^{2}\right\rangle _{(b|I_{\text{A}})}=\left\langle \left(b^{\dagger}\mathcal{A_{\text{B}}}\right)^{2}\right\rangle _{(b|I_{\text{A}})}\nonumber \\
 & = & \mathcal{A_{\text{B}}^{\dagger}}\left\langle bb^{\dagger}\right\rangle _{(b|I_{\text{A}})}\mathcal{A_{\text{B}}}=\mathcal{A_{\text{B}}^{\dagger}}G\mathcal{A_{\text{B}}},\\
\!\!\!\!\!\!\!\!\text{II} & := & \left\langle \left\langle b(s)\right\rangle _{(s|I_{\text{B}})}^{(w_{*})}\left\langle b(s)\right\rangle _{(s|I_{\text{A}})}^{(w_{*})}\right\rangle _{(b|I_{\text{A}})}\nonumber \\
 & = & \mathcal{A_{\text{B}}^{\dagger}}\left\langle bb^{\dagger}\right\rangle _{(b|I_{\text{A}})}\mathcal{A_{\text{A}}}=\mathcal{A_{\text{B}}^{\dagger}}G\mathcal{A_{\text{A}}},\\
\!\!\!\!\!\!\!\!\text{III} & := & \left\langle \left\langle b^{2}(s)\right\rangle _{(s|I_{\text{B}})}^{(w_{*})}\right\rangle _{(b|I_{\text{A}})}=A_{\text{B}}^{\dagger}\widehat{G},
\end{eqnarray}
where we introduced the notation $\widehat{G}(s):=G(s,s)$ for the
diagonal of an operator. The first two of these terms appear in Eq.\ \ref{eq:u_av_av}
in the combination
\begin{eqnarray}
\text{I}-2\,\text{II} & = & \mathcal{A_{\text{B}}^{\dagger}}G\mathcal{A_{\text{B}}}-2\mathcal{A_{\text{B}}^{\dagger}}G\mathcal{A_{\text{A}}}\\
 & = & \left(\mathcal{\mathcal{A_{\text{A}}}-A_{\text{B}}}\right)^{\dagger}G\left(\mathcal{\mathcal{A_{\text{A}}}-A_{\text{B}}}\right)+\mathcal{A_{\text{A}}^{\dagger}}G\mathcal{A_{\text{A}}}\nonumber 
\end{eqnarray}

Inserting this as well as $\text{III}$ into Eq.\ \ref{eq:u_av_av}
gives
\begin{eqnarray}
u(I_{\text{B}}) & = & V(I_{\text{A}})-\frac{w_{I_{\text{A}}}}{2}\times\nonumber \\
 &  & \left[\left(\mathcal{\mathcal{A_{\text{A}}}-A_{\text{B}}}\right)^{\dagger}G\left(\mathcal{\mathcal{A_{\text{A}}}-A_{\text{B}}}\right)+\mathcal{A_{\text{A}}^{\dagger}}G\mathcal{A_{\text{A}}+A_{\text{B}}^{\dagger}\widehat{G}}\right]\nonumber \\
 & = & -\frac{w_{I_{\text{A}}}}{2}\left[\left(\mathcal{\mathcal{A_{\text{A}}}-A_{\text{B}}}\right)^{\dagger}G\left(\mathcal{\mathcal{A_{\text{A}}}-A_{\text{B}}}\right)+\mathcal{A}_{\text{B}}^{\dagger}\widehat{G}\right]\nonumber \\
 &  & +\text{const}(I_{\text{A}}),\label{eq:biased-utility}
\end{eqnarray}
which is maximal for $\mathcal{A_{\text{B}}}=\mathcal{A_{\text{A}}}$
in case $\widehat{G}(s)=b_{\text{A}}^{2}(s)+D(s,s)$ is independent
of $s$, as then $\mathcal{A}_{\text{B}}^{\dagger}\widehat{G}=\widehat{G}(s_{*})$,
with $s_{*}\in S$ arbitrary, is independent of $I_{\text{B}}$. This
happens, for example, in case Alice's uncertainty (covariance) on
Bob's optimal action is the same for all situations. In this case,
Alice optimally communicates properly, with $I_{\text{B}}=I{}_{\text{A}}$,
as this implies $\mathcal{A_{\text{B}}}=\mathcal{A_{\text{A}}}$,
which extremizes the utility.

Otherwise, in case $\widehat{G}(s)$ depends on $s$, her optimal
communication, which she gets by maximizing $u$ w.r.t.\ $\mathcal{A}_{\text{B}}$,
would be such that
\begin{eqnarray}
\mathcal{A_{\text{B}}} & = & \mathcal{A_{\text{A}}}-G^{-1}\frac{\delta\mathcal{A}_{\text{B}}^{\dagger}\widehat{G}}{\delta\mathcal{A}_{\text{B}}}+\lambda,\\
 & = & \mathcal{A_{\text{A}}}-G^{-1}\widehat{G}+\lambda.
\end{eqnarray}
Here $\lambda\in\mathbb{R}$ is a Lagrange multiplier that needs to
be chosen to ensure proper normalization of $\mathcal{A_{\text{B}}}$
via $\boldsymbol{1}^{\dagger}\mathcal{A_{\text{B}}}=\boldsymbol{1}^{\dagger}\mathcal{A_{\text{A}}}=1$,
where we introduced with $\boldsymbol{1}$ a constant unit field,
with $\boldsymbol{1}(s)=1$. This implies $\lambda=\boldsymbol{1}^{\dagger}G^{-1}\widehat{G}/(\boldsymbol{1}^{\dagger}\boldsymbol{1})$.
Thus, the attention Alice optimally would communicate to Bob

\begin{eqnarray}
\mathcal{A_{\text{B}}} & = & \mathcal{A_{\text{A}}}+\left(\frac{\boldsymbol{1}\boldsymbol{1}^{\dagger}}{\boldsymbol{1}^{\dagger}\boldsymbol{1}}-\mathbb{1}\right)\,G^{-1}\widehat{G},\label{eq:bias}
\end{eqnarray}
is her own attention, modified if $G^{-1}\widehat{G}(s)\neq\text{const}$,
if this is possible to her.\footnote{It can turn out that this utility-uncertainty aware optimal attention
function is not larger zero for all $s\in\mathcal{S}$. This would
be an improper attention function, that can not necessarily be communicated
through entropic communication. In this case, we suspect that
\begin{equation}
\mathcal{A_{\text{B}}}=\frac{\left[\mathcal{A_{\text{A}}}+\left(\frac{\boldsymbol{1}\boldsymbol{1}^{\dagger}}{\boldsymbol{1}^{\dagger}\boldsymbol{1}}-\mathbb{1}\right)\,G^{-1}\widehat{G}\right]^{+}}{\boldsymbol{1}^{\dagger}\left[\mathcal{A_{\text{A}}}+\left(\frac{\boldsymbol{1}\boldsymbol{1}^{\dagger}}{\boldsymbol{1}^{\dagger}\boldsymbol{1}}-\mathbb{1}\right)\,G^{-1}\widehat{G}\right]^{+}},
\end{equation}
with $x^{+}:=\text{max}(0,x)=(x+|x|)/2$ being the rectified linear
unit activation function, is then the by Alice optimally communicated
attention function.} For this shift away from her own attention to be desirable, $G$,
her expectations of the second moment of $b$, needs to vary with
$s$ (since the operator $\boldsymbol{1}\boldsymbol{1}^{\dagger}/(\boldsymbol{1}^{\dagger}\boldsymbol{1})-\mathbb{1}$
projects out homogeneous components) as well as not being diagonal
(since then $G^{-1}\widehat{G}=\text{1}$, which is constant w.r.t\ $s$,
and thus projected out by $\boldsymbol{1}\boldsymbol{1}^{\dagger}/(\boldsymbol{1}^{\dagger}\boldsymbol{1})-\mathbb{1}$).

Thus, in case Alice's knowledge about $b$ is homogenous w.r.t.\ $s$,
she wants to properly communicate her attention. If that is not possible,
for example due to a limited set of topics she can use in her communication,
she minimizes a (with $G$ weighted) square distance between her and
Bob's attention function.
\end{document}